\documentclass[12pt,preprint]{aastex}

\newcommand\TC{\tablenotemark{c}}

\newcommand\nd{\nodata}
\newcommand\mcnd{\multicolumn{2}{c}{\nodata}}

\newcommand{\hecs}{\sc{helio10}\upshape}
\newcommand{\tc}{$t^2$}
\newcommand{\tcp}{$\langle t^2\rangle $}

\newcommand{\Hiir}{\ion{H}{2}~region}
\newcommand{\Hiirs}{\ion{H}{2}~regions}
\newcommand{\Hiigs}{\ion{H}{2}~galaxies}

\newcommand{\hel}{\ion{He}{1} line}
\newcommand{\hels}{\ion{He}{1} lines}
\newcommand{\oid}{O$^{++}$/(O$^+$+O$^{++}$)}
\newcommand{\told}{TOL~2146$-$391}
\newcommand{\tolc}{TOL~0357$-$3915} 

\shorttitle{\tc(He$^+$) and \tcp}
\shortauthors{Peimbert, Pe\~na-Guerrero, \& Peimbert}

\received{}
\begin{document}

\title{A CLASSIFICATION OF \ion{H}{2} REGIONS BASED ON OXYGEN AND HELIUM LINES: THE CASES OF \told\ AND \tolc
\footnotemark[1]}

\footnotetext[1]{Based on observations collected at the European Southern 
Observatory, Chile, proposal number ESO 69.C-0203(A).}

\author {Antonio Peimbert\footnotemark[2]}
\email{antonio@astro.unam.mx}

\author{Mar\'ia A. Pe\~na-Guerrero\footnotemark[2]}
\email{guerrero@astro.unam.mx}

\and

\author{Manuel Peimbert\footnotemark[2]}
\email{peimbert@astro.unam.mx}

\footnotetext[2]{Instituto de Astronom\'\i a, Universidad Nacional Aut\'onoma de M\'exico, Apdo. Postal 70-264, M\'exico, C.P. 04510, D.F., Mexico}

\begin{abstract}

We present long slit spectrophotometry of two \Hiigs: \told\ and \tolc. We performed a detailed analysis that involves abundance determinations relaxing the assumption of homogeneous temperature. The temperature inhomogeneities values, \tc, were obtained through two methods: (i) comparing abundances from oxygen recombination lines to abundances from collisionally excited lines and (ii) by using the line intensity ratios of a set of \ion{He}{1} lines together with the \hecs~program. We find that the \hecs~program is a good alternative to obtain a \tc~value in photoionized regions where recombination lines of heavy elements are not available. We have plotted 27 high and low metallicity \Hiirs\ in an oxygen degree of ionization versus \tc~diagram; we find areas populated by \Hiirs\ and areas void of them; the physical characteristics of each area are discussed. In addition, an average \tc~value can be determined for the objects in each area. We propose to use this \tcp~value for the cases where a direct measurement of \tc~cannot be determined.

\end {abstract}

\keywords{galaxies: abundances -- \Hiirs~ galaxies: ISM -- \Hiirs~ regions -- \Hiirs~ ISM: abundances -- \Hiirs~ ISM: individual objects (\told, \tolc)}

\section{Introduction}
\Hiir~chemical abundances provide observational constrains that allow to test models of Galactic chemical evolution through, for example, the determination of radial abundance gradients in spiral galaxies \citep{vila92,zar94}. In addition to this, results derived from nebular studies are used in the field of starburst galaxies which, along with other physical properties such as luminosity or the mass-metallicity relation, allow us to study the chemical enrichment of the Universe \citep[e.g.][]{tre04,per08,liu08,pei11}. It is therefore of great importance for these studies to use accurate abundance determinations.

However, there exists an important problem of inconsistency between the abundances determined with recombination lines (RLs) and those determined with collisionally excited lines (CELs) assuming homogeneous temperature. This is usually refered to as the abundance discrepancy factor (ADF) problem. Arguments in favor of determining abundances from CELs with homogeneous temperature are given by \citet{bre09,erc10,rod10}, among others. Arguments in favor of determining abundances from RLs are given bellow.

There are two main sets of explanations that have been proposed in the literature to reconcile the abundances from CELs and from RLs, thus solving the ADF problem: (i) temperature inhomogeneities within a chemically homogeneous medium, these inhomogeneities would come from shock waves, shadowed regions, advancing ionization fronts, multiple ionizing sources, X-rays, and magnetic reconnection, among others \citep[e.g.][and references therein]{pei06}; and (ii) high metallicity inclusion, which also imply temperature inhomogeneities, these inclusions would be surviving gas from supernova remnants that has not fully mixed with the interstellar medium \citep[e.g.][and references therein]{tsa05,sta07}. Both sets of explanations indicate underabundant results when using CELs and homogeneous temperature to determine abundances. One example of temperature inhomogeneities within a chemically homogeneous medium is given in \citet{pea03}, where a careful study of 30 Doradus was done assuming a chemically homogeneous medium and the obtained abundances were 0.21 dex higher than those obtained using CELs and homogeneous temperature. An example of high metallicity inclusions is given by \citet{tsa05}, whom created a model for 30 Doradus including chemical inhomogeneities that resulted in abundances 0.09 dex lower than those obtained from RLs and 0.12 dex higher than abundances obtained from CELs and homogeneous temperature. 
 
In favor of the abundances derived from RLs is the consistency found in many studies of the chemical evolution of our own Galaxy. Abundances derived from \Hiirs\ considering the O depletion into dust grains and thermal inhomogeneities \citep{car11,pei11} are consistent with: (i) the protosolar abundances corrected by 4.5 Gyr of chemical evolution \citep{asp09,car11}, (ii) the abundances of young F and G stars of the solar vicinity \citep{bens06}, and (iii) the abundances of O and B stars of the Orion region \citep{pry08,sim11}. Throughout this paper we will assume a chemically homogeneous medium and that the abundances are given by RLs or by CELs and considering the presence of thermal inhomogeneitites.

\citet{pei67} first recognized the possibility of including thermal inhomogeneities in studies of photoionized regions; later on, \citet{pei69} studied the magnitude of the effect of such thermal inhomogeneities on the determination of chemical abundances. These two articles introduced the formalism of temperature inhomogeneities (\tc) to determine chemical abundances; in them, it is found that \Hiir~abundances obtained through the traditional method ---which involves assuming a homogeneous temperature throughout the whole volume of the object, usually obtained from the [\ion{O}{3}] line ratio $\lambda\lambda$ 4363/(4959+5007)--- were underestimated by a factor of approximately 2.5 (although, the exact \tc~value is object dependent and its effect on the abundances is also temperature dependent).

With the availability of large telescopes, it has been possible to use the formalism of \tc~to study the ADF. Several studies have shown that ADF typical values are in the 1.5$\,$-$\,$3 range for most \Hiirs~\citep[e.g.][]{pei93,pea05,gara07,pei07,est09}, and in the 1.5$\,$-$\,$5 range for most Planetary Nebulae \citep[e.g.][]{liu93,pei95,sta01,liu06,pei06}. Since CELs lead to underestimate abundances when ignoring the presence of temperature inhomogeneities, we consider that abundances determined from RLs are more reliable.

For chemically homogeneous objects, it is possible to derive accurate abundances from RL ratios (that are independent of \tc), as well as from CELs adopting a \tc~value. There are several ways to determine \tc~values \citep[see][]{pei67,pei11}. The two most frequently used are: (i)  from the comparison of abundances derived from [\ion{O}{3}] lines with those derived from \ion{O}{2} lines (this basically indicates what value of \tc~is required for CELs to reproduce the RL abundances) and (ii) from the comparison of the temperature of forbidden lines to that of the Balmer continuum (this can give accurate O/H abundances without measuring oxygen RLs). Heavy element RLs are particularly faint in low metallicity \Hiirs~such as the ones presented in this paper; moreover, the Balmer discontinuity is smaller and less sensitive at higher temperatures. There is a third method based on \ion{He}{1} lines which has been previously used and which will be further discussed in this paper. For low metallicity objects, where the oxygen RLs are too faint to be measurable and the Balmer discontinuity is very difficult to measure, \ion{He}{1} lines can play a key role in abundance determinations.

Unfortunately, there are some \Hiirs~where a precise value of \tc~cannot be acquired, for instance \Hiigs~with high redshift or low intrinsic brightness. In this scenarios it is important to recognize that, statistically speaking, a good assumption of the value of \tcp~is much better than assuming that \tc=0.00. For those objects, it should be possible to use an average \tc~value obtained from most of the \tc~determinations of \Hiirs~in the literature. In this paper we derive a value of this average for high and low metallicity \Hiirs, \tcp.

In Sections \ref{obs}~and \ref{Icorr}~we present the observations of \told\ and \tolc\ as well as a description of the reduction procedure. The characteristics of the code \hecs\ are presented in Section \ref{He10}. In Section \ref{phys}~we derive temperatures and densities, as well as the available \tc~values. The ionic abundance determinations based on the ratios of CELs to RLs, and based on RLs alone are presented in Section \ref{IonaAb}. In Section \ref{totAbs} we determine the total abundances. In Section \ref{disc}  we present a review of the \tc~values in the literature and recommend an average \tc~value for those objects where a particular \tc~is not available. Finally, in Section \ref{conc} we present the summary and conclusions.

\section{Observations}\label{obs}
The two galaxies presented in this paper were observed in order to increase the number of low metallicity objects with high S/N. Objects of this kind are very useful to study the chemical evolution of galaxies with low SFR, to determine the primordial helium abundance, and to calibrate the lower branch of Pagel's method to determine oxygen abundaces.

\begin{figure}
\includegraphics[angle=0,scale=0.4]{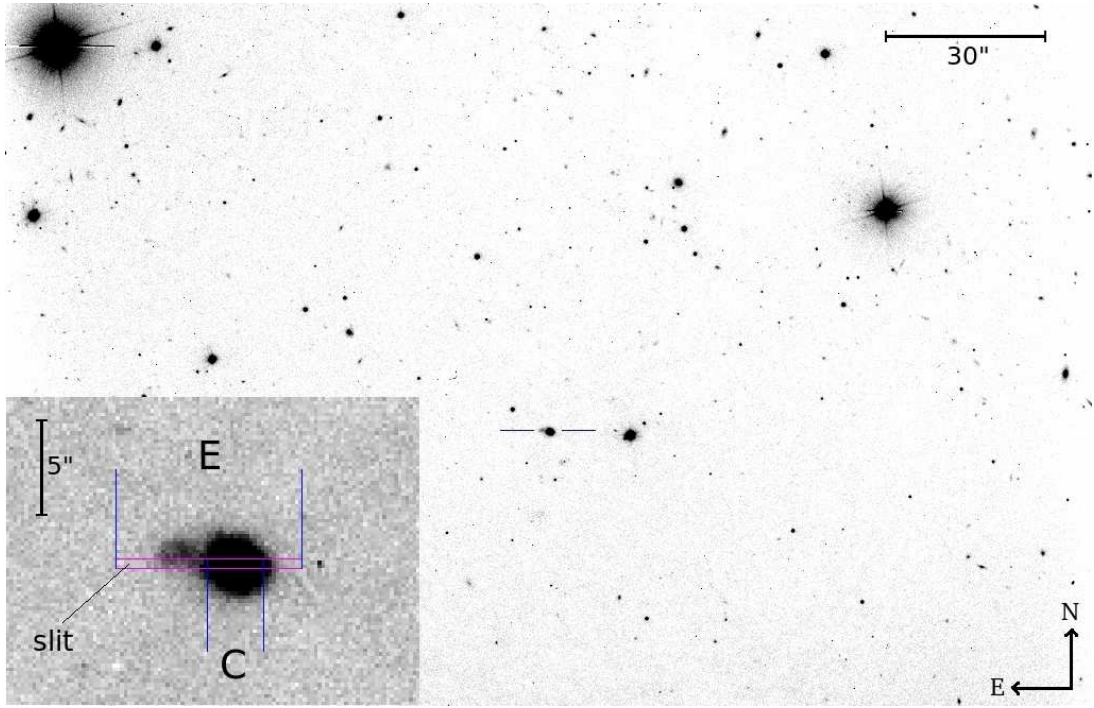}\\
(a)\\
\\
\includegraphics[angle=0,scale=0.4]{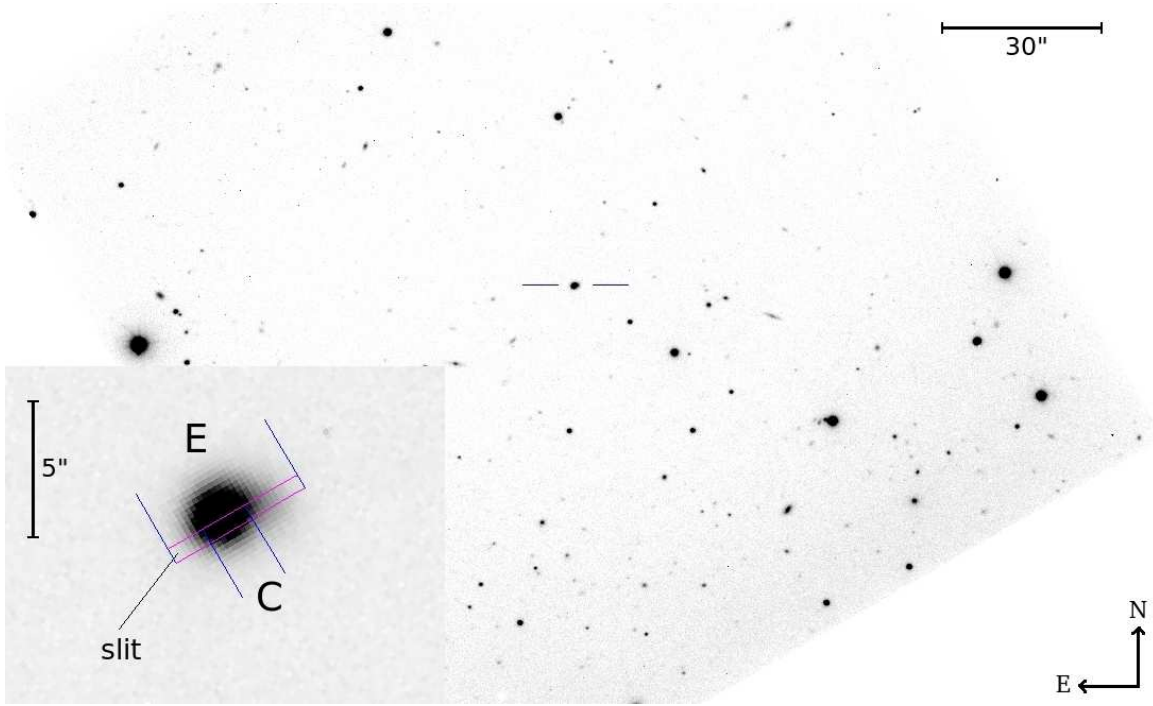}\\
(b)\\
\caption[f2.eps]{
(1a) VLT images of \told\ and (1b) of \tolc. These objects are located at $\alpha=21^h49^m48.2^s$ and $\delta=-38^{\circ}54'08.6''$ and $\alpha=03^h59^m08.9^s$, $\delta=-39^{\circ}06'23.0"$ (J2000.0), respectively. Their redshifts are $z$=0.0295 for \told\ and $z$=0.0744 for \tolc. 
\label{tols}}
\end{figure}

The long slit spectrophotometry presented in this work was obtained at the VLT Melipal facility in Chile, with the Focal Reducer Low Dispersion Spectrograph 1, FORS1. Three grism settings where used: GRIS-600B+12, GRIS-600R+14 with filter GG435, and GRIS-300V with filter GG375 (see Table~\ref{tobs}). Observations were made on September 10$^{\rm th}$ and 11$^{\rm th}$ of 2002 for \told\ ($\alpha=21^h49^m48.2^s$, $\delta=-38^{\circ}54'08.6"$) and \tolc\ ($\alpha=03^h59^m08.9^s$, $\delta=-39^{\circ}06'23.0"$), respectively (see Figure \ref{tols}).

The slit was 0.51" wide and 410" long for both \Hiigs, \told\ and \tolc. The slit was oriented with a position angle of 90$^{\circ}$ to observe \told, and 60$^{\circ}$ for \tolc. The linear atmospheric dispersion corrector, was used to keep the same observed region within the slit regardless of the air mass value. The wavelength coverage and the resolutions for the emission lines observed with each grism are given in Table \ref{tobs}. The average seeing during the observations amounted to 0.74" for \told\ and 0.88" for \tolc.

We took two extractions from each object, Extraction C stands for the core of each \Hiir, while Extraction E stands for each extended \Hiir. In \told\ Extractions C and E had 15 pixels (3.0'') and 49 pixels (9.8''), respectively; while for \tolc\ Extractions C and E had 9 (1.8'') and 27 pixels (5.4''), respectively. Extractions C are centered at the brightest region while Extractions E are slightly displaced in order to include most of the emission (see Figure \ref{tols}). For \told, Extraction C included 94\% of the flux included in Extraction E, and for \tolc, Extraction C included 88\% of the total flux of Extraction E. We decided to present the analysis of both extractions of both objects for several reasons: (i) we were trying to obtain as much information as possible with the highest precision available, (ii) we were looking to see if there were drastic differences between the central and the outer parts of each extraction (notice the plots present in the inserts), (iii) we were looking for the determination with the best S/N available, and (iv) both analysis were completely independed and helped us to better estimate the errors in the line intensity determinations. 

\begin{figure}
\begin{center}
\includegraphics[angle=0,scale=0.8]{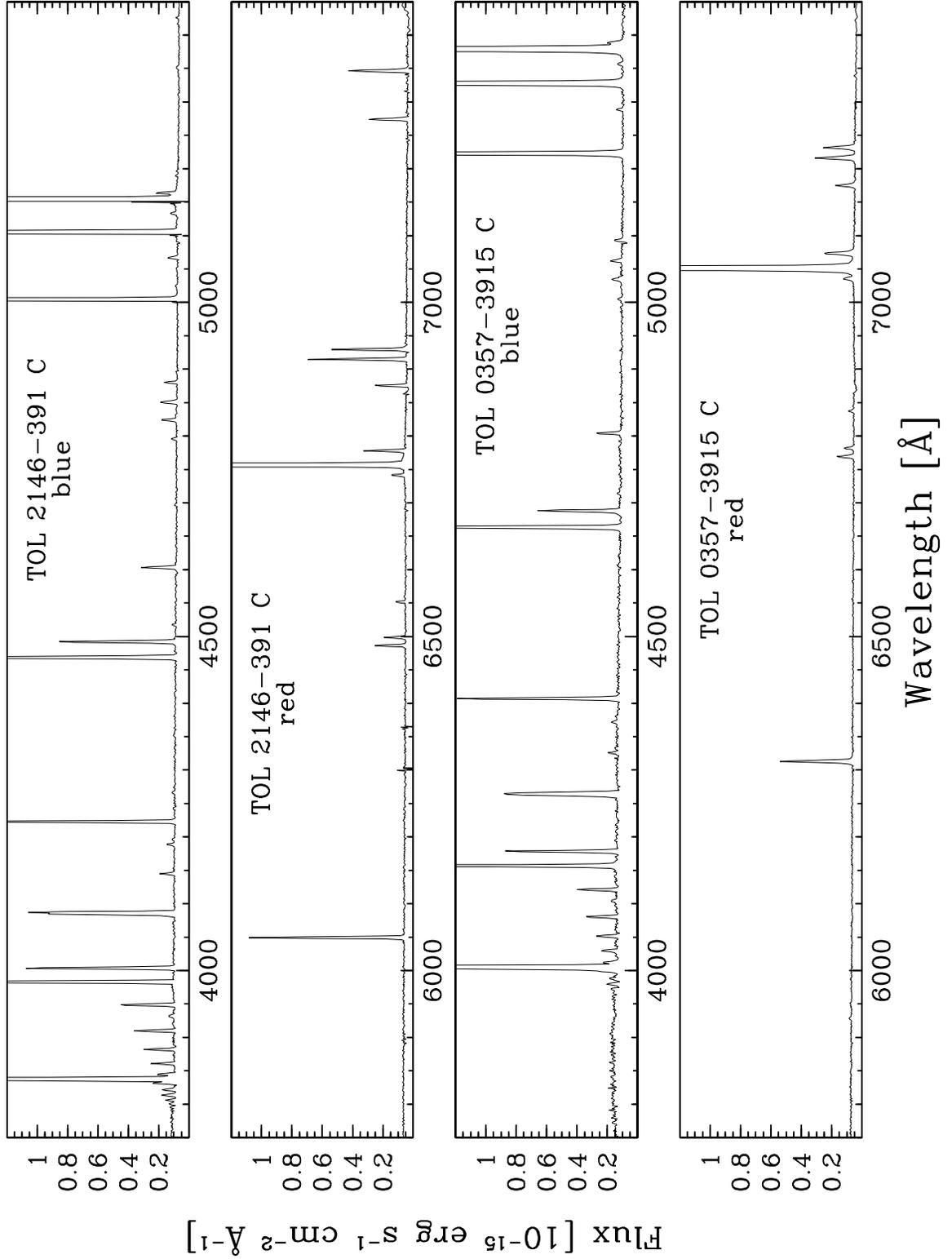}
\caption[f2.eps]{
Plot of most of the observed wavelength range for both high resolution spectra of the core extractions. \label{especs}}
\end{center}
\end{figure}

The spectra were reduced using IRAF\footnotemark{}, following the standard procedure of bias subtraction, aperture extraction, flatfielding, wavelength calibration, and flux calibration. For flux calibration the standard stars LTT~2415, LTT~7389, LTT~7987, and EG~21 were used \citep{ham92,ham94}. The observed spectra are presented in Figure \ref{especs}.

\footnotetext{IRAF is distributed by the National Optical Astronomy Observatory, which is operated by AURA, Inc., under cooperative agreement with the National Science Foundation.}

\section{Line Intensities and Reddening Correction}\label{Icorr}
The \begin{tt}splot\end{tt}~task of the IRAF package was used for line measurements. Line intensities were obtained by integrating the flux in the line between two limits over the local continuum estimated by eye. When line blending was present, a multiple Gaussian profile was used to determine the line intensities. 

We used the measurements of the Balmer decrement of the low resolution spectra to tie the blue and red high resolution spectra. The emission line intensities of all extractions in \told\ and in \tolc\ are presented in Tables \ref{tlines}~and \ref{tlines2}, respectively. 

To determine the reddening correction we compared the theoretical Balmer line intensity ratios with the observed ones. A proper determination of the reddening correction is always desired, in this work it is required to allow \hecs\ to be able to extract all the possible information from the \ion{He}{1} lines that span from $\lambda\lambda$ 3634 to 7281.

The observed Balmer line intensities have to be corrected for reddening, underlying absorption, and collisional excitation before they can be compared to the theoretical recombination intensities; also both ---the theoretical Balmer recombination intensities and the collisional Balmer excitation intensities--- depend on the temperature; and to obtain estimates of the temperature it is necessary to use dereddened line intensity ratios. This means that all four quantities need to be fitted self-consistently.

A first approximation to the correction for the reddening coefficients, $C$(H$\beta$)s, is obtained by ignoring underlying absorption and collisional excitation, adopting $T_e=$10,000 K, and fitting the observed and theoretical $I$(H$\alpha$)/$I$(H$\beta$) ratios. This reddening correction is applied to the [\ion{O}{3}] lines to obtain $T_e$[\ion{O}{3}]. This approximation will be enough to determine (i) the theoretical intensities of the Balmer lines ---that are only weakly dependent on the temperature---, and (ii) the contribution of collisional excitation to the Balmer line intensities ---that is small \citep{pei07}. Finally, the reddening correction and underlying absorption were fitted simultaneously taking advantage of the fact that reddening is more important for H$\alpha$, H$\beta$, and H$\gamma$, while underlying absorption is more important for the weaker bluer Balmer lines. (Notice that the contribution of \ion{He}{2} $\lambda$ 4859 to H$\beta$ is less than one part in a thousand).

For both extractions of \tolc, the best fit to the four brightest Balmer lines yielded an H$\beta$ greater than 100 (102.20 for Extraction C, and 100.80 for Extraction E), which is within the observational errors presented in Table \ref{tlines2}. We took the unconventional decision of choosing this H$\beta$ value to give the proper weight to these four Balmer lines. 

The theoretical Balmer recombination intensity ratios were obtained using the program \begin{tt}INTRAT\end{tt}~ \citep{sto95}~assuming $T_e$=13,415 K and $n_e$=140 cm$^{-3}$ for \told, and $T_e$=12,570 K and $n_e$=120 cm$^{-3}$ for \tolc\ (forbidden line intensity ratios used to determine densities are not dependent on any of the parameters we are adjusting in this section). \begin{tt}INTRAT\end{tt}~stands for INTensity RATios of hydrogenic recombination lines for specified transitions.

The fraction of the Balmer line intensities due to collisional excitation from the ground state were estimated from interpolations to the models presented by \citet{pei07}. This correction is not commonly used because it is very small for objects cooler than 12,000~K as well as for ionized regions that are density bounded. However, it is relevant to both, the proper determination of $C$(H$\beta$) and $I_{rec}$(H$\beta$), for these two objects; contributing with 3.8\% and 3.2\% for the $I$(H$\alpha$) of \told\ and \tolc, respectively. We normalized to $I_{rec}$(H$\beta$) because most of the formulas that compare line intensities to that of H$\beta$, implicitely assume that $I$(H$\beta$) represents the recombination cascade of the nebular component. Due to the energy level distribution, collisional excitation from the ground level is negligible for the observed permitted lines of all the other ions.

The reddening coefficients, were determined by assuming the extinction law of \citet{sea79}, $f(\lambda)$. The values found for $C$(H$\beta$) as well as for the assumed $f(\lambda)$ are presented in Tables \ref{tlines}~and \ref{tlines2} for \told\ and \tolc, respectively.

The correction due to the underlying absorption equivalent widths for the Balmer lines was done using a stellar spectra template normalized to $EW_{abs}$(H$\beta$). The template is presented in Table \ref{tEWs} and the normalizations used are presented in Tables \ref{tlines} and \ref{tlines2}. The template was generated based on the low metallicity instantaneous burst models from \citet{gon99}. 
These absorption line ratios do not change considerably during the first 5 Myr, the numbers presented in Table \ref{tEWs} represent an average of the ratios of the models between 0 and 5 Myr. 
To complete the template we require H$\alpha$ and H7, M. Cervi\~no kindly ran for us some models with the same code as R. Gon\'zalez-Delgado and collaborators, and provided us with the $EW_{abs}$ for many lines required in this template.
The high-n Balmer lines are too close to be separated in the spectra provided by this code; alternatively, it is possible to estimate the $EW_{abs}$ from a deep echelle observation of a young \Hiir. For these, we used the data of 30 Doradus obtained by \citet{pea03}, where we estimated the $EW_{abs}$(H$\,$high-n)/$EW_{abs}$(H$\beta$) required to adjust the observed Balmer lines to the theoretical ones.
The $EW_{abs}$(H$\beta$) values we adopted do not correspond to any particular age, but do correspond to the intensities that best fit the high n Balmer lines to H$\beta$ ratios.
Forbidden lines do not have underlying absorption, however, permitted lines do; underlying absorption equivalent widths used for the \ion{He}{1} lines included in the template come from Gonz\'alez-Delgado and colaborators as well as the additional models ran by Cervi\~no and are also presented in Table \ref{tEWs}. The corrections to the line intensities due to underlying absorption of \ion{C}{2}, \ion{N}{2}, and \ion{O}{2} are expected to be negligible, more than an order of magnitude smaller than our observational uncertainties, and were not considered.

Column (1) in Tables \ref{tlines}~and \ref{tlines2}~presents the adopted laboratory wavelength, $\lambda$, and column (2) the identification for each line. Column (3) in both tables presents the extinction law value used for each line \citep{sea79}.  In both tables, columns (4-6) represent the data of the core extractions, Extraction C, while columns (7-9) represent the data of the extended extractions, Extraction E. Columns (4) and (7) include the observed flux relative to $F$(H$\beta$)=100.00, $F(\lambda)$; columns (5) and (8) include the flux corrected for reddening (as well as underlying absorption ---for \ion{H}{1} and \ion{He}{1} lines--- and collisional excitation ---for \ion{H}{1} lines) relative to $I_{rec}$(H$\beta$)=100.00, $I$($\lambda$); finally columns (6) and (9) include the error percentage associated with those intensities, which already include all the errors in $C$(H$\beta$), $EW_{abs}$(H$\beta$), and collisional excitation of the Balmer lines.

\section{\hecs}\label{He10}
Recombination lines are not usually used to determine physical quantities of the ionized regions: the intensities of all \ion{H}{1} lines are approximately proportional to $n_e n$(H$^+) T_e^{-0.9}$ so their ratios are nearly independent of $n_e$ and $T_e$; this is the reason they are ideal to determine the reddening correction. To determine $T_e$ using the Balmer discontinuity it is necessary to have an exquisitely accurate tailored made model of the stellar spectra, since about $80\%$ of the observed continuum is stellar and the Balmer discontinuity shows up in absorption \citep[at very high densities it is possible to use high order Balmer lines to estimate $n_e$, e.g.][]{liu00,pei04}. \ion{He}{2} lines are not more useful than \ion{H}{1} lines to determine $n_e$ or $T_e$, plus, they are generally fainter. On the other hand, RLs of heavy elements could theoretically provide information to derive the physical conditions, but are usually too faint to do much with them \citep[although very deep echelle observations of the \ion{O}{2} multiplet 1 lines have been shown to be useful for $n_e$ determinations, e.g.][]{rui03,pea05}.

\ion{He}{1} lines are a few orders of magnitude stronger than heavy element recombination lines and have a less homogeneous (more complicated) temperature and density dependence than RLs from hydrogenoid ions making \ion{He}{1} lines the best RLs to determine physical conditions. The intensity of each \ion{He}{1} line is proportional to $n_e n($H$^+) T_e^{-\alpha}$, however, each line has a slightly different $\alpha$ ---for the brightest \ion{He}{1} lines, $\alpha$ is in the 0.7 to 1.1 range--- making it possible to determine the temperature from a ratio of \ion{He}{1} lines. Also, adding information and complexity to the determinations, the metastable level 2$^3$S produces an additional dependence on density and optical depth, $\tau$(2$^3$S). The 2$^3$S level is stable for 2.5 hours before radiatively decaying, thus providing enough excited atoms to produce an appreciable effect on (i) the optical depth to the n$^3$P-2$^3$S \ion{He}{1} lines; these, in turn, affect many of the triplet lines (\ion{He}{1} $\lambda\lambda$ 3889 and 7065 are the most affected optical lines); as well as (ii) an appreciable target for collisional excitations which in turn enhances all the optical \ion{He}{1} lines (again, the most affected are \ion{He}{1} $\lambda\lambda$ 3889 and 7065). In the end, each triplet \ion{He}{1} line has a unique dependence on $n_e$, $T_e$, and $\tau$(2$^3$S); while singlet \ion{He}{1} lines depend on $n_e$ and $T_e$ (singlet lines are not affected by the opacity produced by the metastability of the 2$^3$S level, they only depend weakly on the opacity produced by the ---much smaller--- metastability of the 2$^1$S level). Even with the additional complications of having one more variable, triplets tend to be better than singlets because they have approximately 3 times more intensity and have a stronger dependence on $n_e$.

Properly measured \ion{He}{1} line intensities of the photoionized material (i.e. lines that have been properly corrected for the reddening produced by dust grains and for the underlying absorption present in the stellar spectra), can be used to determine the average physical conditions of such gas parcel. Theoretically, one needs at least four \ion{He}{1} lines ---three independent \ion{He}{1} line ratios--- to be able to solve for all the unknowns (i.e. $n_e$, $T_e$, and $\tau$(2$^3$S) along with the $n$(He$^+$)/$n$(H$^+$) ratio); in practice, more lines are better since dependences on $n_e$, $T_e$, and $\tau$(2$^3$S) is weak for most line ratios.

\hecs~is an extension of the maximum likelihood method used in previous works to determine \tc(He$^+$) \citep[see][]{pei00,pea02,pea03,est04,pea05}. This version of the code can determine $T_e$(\ion{He}{1}) using only \ion{He}{1} line intensities. The code also yields $T_0$ and $t^2$ (the formal definitions are presented in Section \ref{Tinhom}) by combining the information of the \ion{He}{1} lines with the information of the [\ion{O}{2}] (7325/3727) and [\ion{O}{3}] (4363/5007) line ratios.

The effective recombination coefficients for the \ion{H}{1} and \ion{He}{1} lines were those given by \citet{sto95} for H, and by \citet{ben99} and \citet{por07} for He. The collisional contribution was estimated from \citet{saw93} and \citet{kin95}. The optical depth effects in the triplets were estimated from calculations made by \citet{ben02}. The collisional strengths come from \citet{mcl93} for [\ion{O}{2}] and \citet{len94} for [\ion{O}{3}], while the transition probabilities come from \citet{wie96}.

In its simplest mode \hecs\ uses up to 20 \ion{He}{1}/H$\beta$ ratios with their uncertainties to search for the most likely values for $T_e$(\ion{He}{1}), $n_e$(\ion{He}{1}), $\tau$(2$^3$S), and $n$(He$^+$)/$n$(H$^+$). It firsts assumes a certain set of values and, using the relevant atomic data, estimates the theoretical \ion{He}{1}/H$\beta$ ratios. Then, it compares the observed ratios to the theoretical ones determining
\begin{equation}
\chi^2=\sum\limits_{\lambda}
{\left\{ \left[{I({\lambda}) \over I({\rm H\beta})}\right]_{obs} - 
\left[{I({\lambda}) \over I({\rm H\beta})}\right]_{theo} \right\}^2
\over
\sigma_{I(\lambda)}^2}.
\end{equation}
Using multidimensional optimization algorithms, \hecs~then searches the parameter space looking for the minimum $\chi^2$ value, $\chi^2_{min}$; and, finally, it determines the available parameter space within $\chi^2_{min} \leq \chi^2 \leq \chi^2_{min}+1$ to estimate the 1$\sigma$ error bars of each one of the four variables.

Additional constraints can be included to better determine the physical conditions of the nebula on any or all of the variables; any available determination independent of the \ion{He}{1} lines can be used as an input (with the corresponding uncertainty). When the solution given by the \ion{He}{1} lines alone is very uncertain, an additional constraint directly restricts one variable and indirectly restricts all the others; when the solution given by the \ion{He}{1} lines alone has uncertainties of a similar magnitude as those given by the additional constraint, the output for that variable becomes a weighted average while the others become slightly better restricted (when using these additional constraints the corresponding variables will have an output that does not necessarily match the partial input).

In general the temperatures derived fom the \hel\ data are smaller than those derived from forbidden line data. When combining \ion{He}{1} data with forbidden line data, to determine self-consistent physical conditions, instead of searching for $T_e$(\ion{He}{1}) \hecs\ searches for $T_0$ and $t^2$ (as well as for $n_e$(\ion{He}{1}), $\tau$(2$^3$S), and $n$(He$^+$)/$n$(H$^+$)); this $T_0$ and $t^2$ together with a specific shape define a distribution of temperatures. \hecs\ simulates five types of temperature distributions: gaussian, square, symmetric triangle, left triangle, and right triangle. For a certain set of \ion{He}{1} lines, the exact determination of $t^2$ depends on the shape of the distribution: right triangles require slightly larger $t^2$ than left triangles, and square distributions require slightly larger $t^2$ values than symmetric triangle distributions, gaussian distributions lie in the middle. The dispersion between the distributions is less than 10\% on the $t^2$ values except for the most extreme cases. To be able to distinguish between these distributions using \ion{He}{1} lines alone, the uncertainties in the intensity determinations would be required to be at least two orders of magnitude smaller than the observed ones; the recommended shape of the distribution depends on the mechanism responsible for the thermal inhomogeneities. For this work we will present the results obtained using gaussian distributions. \hecs\ does not model density variations, but the models are robust for \Hiirs\ because their densities usually are lower than the critical density of the 2$^3$S level, $n_{crit} \approx 3,000\, {\rm cm}^{-3}$ (critical densities for the [\ion{O}{2}] (7325/3727) and [\ion{O}{3}] (4363/5007) ratios are much higher).

The practical limitation of the \hecs\ program when determining $t^2$, is the number and signal-to-noise (S/N) of the \ion{He}{1} lines it requires. For a fair $t^2$ determination, with the error in the 0.030-0.040 range, at least 8 \ion{He}{1} lines are required, 5 of them with a S/N greater than 20. For a precise $t^2$ determination, with an error smaller than 0.020, at least 5 \ion{He}{1} lines with a S/N greater than 35 are required. For \told, the S/N of the fifth best \ion{He}{1} line are in the 20 to 25 range, while for \tolc\ they are in the 10 to 25 range.

When working with \hecs, the five most accessible/useful lines usually are: $\lambda\lambda$ 3889, 4471, 5876, 6678, and 7065; the lack of any one of these lines strongly diminishes the quality of the determinations. These lines are about 300 (for $\lambda$ 5876) to 50 (for $\lambda$ 7065) times brighter than oxygen RLs in oxygen-poor \Hiirs\ like the ones presented in this paper (12+log(O/H)$\approx$8.15).

Due to the similar ionization potentials of He$^+$ and O$^{++}$, the $t^2$ derived from \ion{He}{1} lines can be used to determine abundances considering thermal inhomogeneities in \Hiirs\ with high O$^{++}$/O ratios.

\section{Physical Conditions}\label{phys}

\subsection{Temperatures and Densities from CELs}
Temperature and density measurements determined from CELs are presented in Table \ref{tdat}. All temperatures of heavy elements as well as densities from  [\ion{O}{2}], [\ion{S}{2}], and [\ion{Cl}{3}] were determined fitting the line intensities in Tables \ref{tlines}~and \ref{tlines2}~to the \begin{tt}temden\end{tt} task in IRAF; which models a five-, six-, or eight-level ion to derive the physical conditions.

The [\ion{Fe}{3}] density was estimated using the theoretical intensity determinations presented by \citet{kee01}. The tables presented in their paper show that (i) for $n_e\le300\,$cm$^{-3}$, the two brightest optical [\ion{Fe}{3}] lines are $\lambda\lambda$ 4658 and 4986, and (ii) their ratio is strongly density dependent, going from $I(4986)/I(4658)\approx1.08$ at $n_e=100\,$cm$^{-3}$ to $I(4986)/I(4658)\approx0.05$ at $n_e=3,000\,$cm$^{-3}$; making this ratio an excellent density indicator ---specially since, at low densities, it is much more sensitive than the [\ion{O}{2}], [\ion{S}{2}], and [\ion{Cl}{3}] density determinations. Unfortunately, the computations presented by Keenan and collaborators stop at $n_e=100\,$cm$^{-3}$, however, extrapolations to slightly lower values should be accurate. 

For the purpose of observations with spectral resolution in the $1,000<{\Delta\lambda/\lambda}<10,000$ range, such as the ones presented in this paper, it is better to try to derive the density using the [\ion{Fe}{3}] $I(4986+4987)/I(4658)$ ratio. From the data of \citet{kee01}, we fitted:
\begin{equation}
\log\left({I(4986+4987) \over I(4658)}\right)=0.05+0.25(\log T_e-4)-0.66(\log n_e-2)+0.18(\log T_e-4)(\log n_e-2),
\end{equation}
which adjusts the calculated ratios for $7,000\,$K$<T_e<20,000\,$K and $100\,$cm$^{-3} < n_e < 300\,$cm$^{-3}$ to better than 2\% and for $100\,$cm$^{-3} < n_e < 1000\,$cm$^{-3}$ to better than 5\%; we expect this ratio to be a good estimate (better than 5\%) for $30\,$cm$^{-3} < n_e < 1000\,$cm$^{-3}$. From this fit, the line intensities, and $T_e$ we can estimate:
\begin{equation}
\log n_e=2-\left[{\log\left({I(4986+4987) \over I(4658)}\right)-0.05-0.25(\log T_e-4)\over0.66-0.18(\log T_e-4)}\right];
\end{equation}
the determinations of $n_e$[\ion{Fe}{3}] presented in Table \ref{tdat} were obtained using this equation.

\subsection{Temperatures and Densities from RLs}
The only temperatures and densities available from RLs are those of \ion{He}{1} that were obtained using the program \hecs~(see Section \ref{He10}); they are presented in Table \ref{tdat}. It can be seen that for \told\ the RL temperatures are 2,250 K lower than those from CELs, showing that this \Hiir~cannot be reproduced by using simple (homogeneous temperature) models. For the case of \tolc, the temperature difference is only 750 K and the error bars are larger; however, there are strong reasons to think that the true $T_e$(\ion{He}{1}) is even lower and not closer to $T_e$[\ion{O}{3}] implying a non negligible correction to the chemical abundances (see section \ref{Tinhom}).

Other RL determinations such as $n_e$ from the multiplet 1 of \ion{O}{2} or $n_e$ from high order Balmer lines are not available due to the quality of the observations. Also, $T_e$(Bac) is unavailable because the nebular Balmer continuum discontinuity has been distorted by the presence of the stellar Balmer continuum in absorption.

\subsection{Temperature Inhomogeneities}\label{Tinhom}
In order to account for an inhomogeneous temperature structure throughout the objects, we used the \tc~formalism developed by \citet{pei67}. This formalism is designed to study a photoionized object that has both thermal structure and ionization structure (as are all real astronomical objects); each ionic species has its own average temperature and its own value of thermal inhomogeneities:

\begin{equation} T_0(ion) \equiv \frac{\int T_e(\textbf{r})n_e(\textbf{r})n_{ion}(\textbf{r}) dV}{\int n_e(\textbf{r})n_{ion}(\textbf{r}) dV}
\label{T0}, \end{equation}

\begin{equation} t^2(ion) \equiv \frac{\int (T_e-T_0)^2 n_e(\textbf{r})n_{ion}(\textbf{r}) dV}{T_0^2 \int n_e(\textbf{r})n_{ion}(\textbf{r}) dV},
\label{t2} \end{equation}
with $n_e$, $n_{ion}$, and $V$ representing the electron density, ion density, and the observed volume, respectively.

We used two independent sets of temperatures to derive $T_0$ and \tc, one set whose intrinsic average weights preferentially the high-temperature regions and one temperature that weights preferentially the low-temperature regions \citep{pei67}. The temperatures that weight preferentially high-temperature regions where $T_e$[\ion{O}{2}] and $T_e$[\ion{O}{3}] for the low and high-ionization zones, respectively. These temperatures can be represented as a function of $T_0$ and \tc~as in \citet{pei04}:

\begin{equation} T_{4363/5007} = T_{0} \Bigg[ \bigg(1 + \frac{t^2}{2} \left( \frac{91300}{T_{0}} -3\right) \Bigg]
\label{TO3}
\end{equation}
and
\begin{equation} T_{7325/3727} = T_{0} \Bigg[ \bigg(1 + \frac{t^2}{2} \left( \frac{97800}{T_{0}} -3\right) \Bigg].
\label{TO2}
\end{equation}
The temperature that weights preferentially the low-temperature region was determined from the \ion{He}{1} lines with good S/N. We used the \hecs~program to find $T_e$(\ion{He}{1}) and \tc(He$^+$); for a detailed description of the program see Section \ref{He10}. The \tc~values we obtained with the \hecs~program for each extraction in both objects were: \tc(He$^+$)=0.096$\pm$0.038 for \told\ C, \tc(He$^+$)=0.098$\pm$0.055 for \told\ E, \tc(He$^+$)=0.028$\pm$0.064 for \tolc\ C, and \tc(He$^+$)=0.004$\pm$0.054 for \tolc\ E.

The \tc~values found for \tolc\ have errors that are too large, they are consistent with large temperature inhomogeneities as well as with no temperature inhomogeneities; therefore, it is not possible to work with these \tc~values with good confidence.

To obtain another value of the thermal inhomogeneities, \tc(O$^{++}$), we also used the temperature derived from the ratio of the RLs of the multiplet 1 of \ion{O}{2} to the CELs of [\ion{O}{3}] as given in \citet{pea05} \citep[see also][]{pea10}:
\begin{equation} T_{{\rm O II/ [O III]}} = T_{4651/5007} = f_1 (T_0,t^2).
\label{TOrec}
\end{equation}
From the oxygen RLs in \told\ we were able to obtain $T_0=$13,765 K and \tc(O$^{++}$)= 0.084$\pm$0.041 for Extraction C, and $T_0=$14,524 K and \tc(O$^{++}$)=0.113$\pm$0.044 for Extraction E. For \tolc\ we did obtain measurements of the oxygen RLs in both extractions, however, the S/N was too small to determine reliable O abundances.

We combined the \tc~values in \told\ to obtain a final value of the thermal inhomogeneities for each extraction; these amounted to \tc=0.091$\pm$0.028 for \told\ C and \tc=0.107$\pm$0.034 for \told\ E.

When a \tc~determination is consistent with zero, but has an error greater than about 0.030, it does not mean that the object has negligible temperature inhomogeneities: it means that the method used to determine \tc~does not have good enough S/N; nonetheless, it is better to use the derived \tc~value to determine abundances than to assume that the temperature is homogeneous, otherwise abundances will be systematically underestimated. For objects where \tc~values are poorly determined, it is even better to use an average \tc~of objects with similar physical conditions. All \tc~values depend on the specific characteristics of the thermal structure of each nebula and, although well determined \tc~values from observations of \Hiirs~range between 0.020 and 0.120, we recommend to use \tcp~only if not enough information is available to obtain a particular \tc~value.

Since there was not enough information in the \hels~nor in the \ion{O}{2} RLs of \tolc\ to obtain a precise determination of \tc, we decided to follow the recipe derived in Section \ref{disc}. Since this object has a value of \oid$\ge0.75$, we adopted \tc$=0.051\pm0.026$.

\section{Ionic Abundances}\label{IonaAb}

\subsection{Ionic Abundances from CELs}\label{IonAbCELs}
Available ionic abundances of both objects are presented in Table~\ref{tceionic}. These were determined using the task \begin{tt}abund\end{tt} of IRAF. For \told\ C and E we used $T_{\rm[OII]}=$14,120 K and 14,140 K, and $T_{\rm[OIII]}=$ 15,800 K and 15,810 K, respectively. For \tolc\ C and E we used $T_{\rm[OII]}=$13,600 K and 13,720 K, and $T_{\rm[OIII]}=$ 14,870 K and 14,770 K, respectively.

Ionic abundances considering thermal inhomogeneities, \tc$\neq$0.00, were obtained using the traditional determinations corrected by the formalism presented by \citet{pei69} \citep[see also][]{pei11}. Abundances assuming a homogeneous temperature, \tc=0.00, are also presented in Table~\ref{tceionic} only for the purpose of comparison with other authors.

\subsection{Ionic Abundances from RLs}\label{IonRLs}

\subsubsection{Helium RLs}\label{HeI}
The He$^+$ abundances were determined with the \hecs~program and the lines presented in Tables \ref{tlines}~and \ref{tlines2}~(see Section \ref{He10}). He$^{++}$ abundances were determined using the H$^+$ and He$^{++}$ recombination coefficients given by \citet{sto95}~in their program \begin{tt}INTRAT\end{tt}, as well as the $\lambda$ 4686 line from Tables \ref{tlines}~and \ref{tlines2}. Ionic He abundances are presented in Table \ref{trionic}.

\subsubsection{Oxygen RLs}\label{IonAbORLs}
Measurements of the RLs of multiplet 1 of \ion{O}{2} in \told\ are presented in Table~\ref{tlines}, and ionic O abundances from RLs are presented in Table \ref{trionic}. We were also able to measure the oxygen RLs in both extractions of \tolc, nonetheless the errors in these measurements are too large to provide useful information. In all cases, we were able to detect only four of the eight lines in the multiplet. Due to the spectral resolution of the observations, those lines were blended into two pairs: $\lambda \lambda$4639$+$42 and $\lambda \lambda$4649$+$51.

For a density of 280 cm$^{-3}$, the sum of $\lambda \lambda$4639$+$42 and $\lambda \lambda$4649$+$51 represents about 69\% of the multiplet \citep{pea05,pea10}. We used equations 3 and 4 of \citet{pea10} to obtain the ionic abundance of the oxygen RLs in \told.

\section{Total Abundances}\label{totAbs}
The ionization correction factors (ICFs) used to obtain total abundances are presented  in Table \ref{tICF}, total gaseous abundances for all available elements are presented in Table~\ref{tabund} for \told, and in Table~\ref{tabund2} for \tolc. Within the errors, abundances in both extractions of \told\ are very similar as well as in \tolc. Observations with higher S/N of these objects would allow to measure a more precise value of the thermal inhomogeneities in both objects.

Total gaseous abundances were calculated using the following equations:
\begin{equation}   \frac{n({\rm N})}{n({\rm H})}={\rm ICF(N)} \times \frac{n({\rm N^{+}})}{n({\rm H^+})},
\label{totN}
\end{equation}
\begin{equation}   \frac{n({\rm O})}{n({\rm H})}={\rm ICF(O)} \times  \frac{n({\rm O^{+}})+n({\rm O^{++}})}{n({\rm H^+})},
\label{totO}
\end{equation}
\begin{equation}
\frac{n({\rm Ne})}{n({\rm H})}={\rm ICF(Ne)}  \times \frac{n({\rm Ne^{++}})}{n({\rm H^+})},
\label{totNe}
\end{equation}
\begin{equation}   \frac{n({\rm Cl})}{n({\rm H})}={\rm ICF(Cl)} \times \frac{n({\rm Cl^{++}})}{n({\rm H^+})},
\label{totCl}
\end{equation}
\begin{equation}   \frac{n({\rm S})}{n({\rm H})}={\rm ICF(S)} \times \frac{n({\rm S^{+}})+n({\rm S^{++}})}{n({\rm H^+})},
\label{totS}
\end{equation}
and
\begin{equation}   \frac{n({\rm Ar})}{n({\rm H})}={\rm ICF(Ar)} \times \frac{n({\rm Ar^{++}})+n({\rm Ar^{+++}})}{n({\rm H^+})}.
\label{totAr}
\end{equation}
The estimated ICF values, as well as the references to the analytical approximations to the ICF values, are presented in Table \ref{tICF}.

To obtain the total O/H abundances, a correction of 0.08$\,$-$\,$0.12 dex should be added to the gaseous O/H abundances due to oxygen depletion into dust \citep{pea10}. Therefore a correction of 0.10 dex is already included in the total O/H abundances presented in this work (see Table \ref{totAbs}).

\section{Discussion}\label{disc}
In this paper we have assumed that the ADFs are caused by the presence of temperature inhomogeneities in a chemically homogeneous medium, other options are mentioned in the introduction. As mentioned in Section \ref{Tinhom}, there was not enough information in the observed lines of \tolc\ to obtain a meaningful \tc\ determination. We decided to search the literature for all the \Hiirs\ with good \tc\ determinations in order to find an average \tc\ value that could be representative for our object. For objects with several \tc\ determinations, we chose only the ones with the smallest error bars; to these objects we added Extraction C of \told. Table \ref{tWMR} presents the 27 object sample we found, 8 of them are Galactic \Hiirs\ and the other 19 are extragalactic \Hiirs. For each object we present the oxygen abundance derived from CELs and \tc=0.000, as well as the oxygen abundance assuming the presence of thermal inhomogeneities including the fraction of oxygen depleted into dust grains; we also present the \tc\ determination, the oxygen ionization degree given by \oid, the [\ion{O}{3}] temperature, and the [\ion{O}{2}] density.  

\begin{figure}
\begin{center}
\includegraphics[angle=0,scale=0.80]{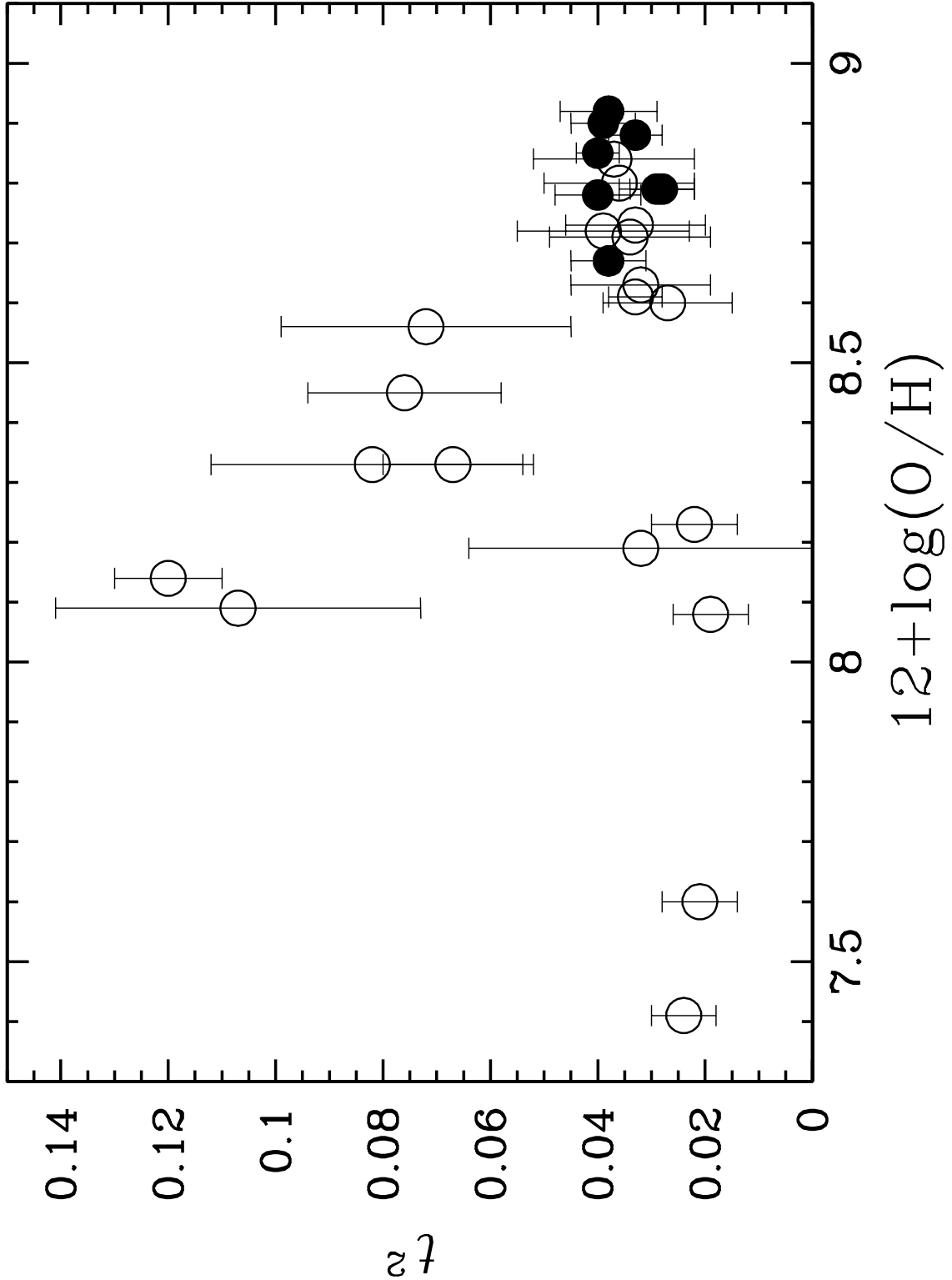}
\caption[f3.eps]{
Plot of 12+log(O/H) values with correction for \tc~and dust depletion of O versus the measured \tc~values. The filled circles are Galactic \Hiirs\ and the open circles are the extragalactic \Hiirs. \label{t2prom}}
\end{center}
\end{figure}

In Figure \ref{t2prom}~we present the O/H versus \tc~diagram for the data presented in Table \ref{tWMR}. Galactic objects are presented by filled circles and extragalactic objects are presented by open circles. For those 5 objects without a quoted error in the \tc\ determination, we adopted an error bar of 40\% of the \tc\ measured value. From this figure it can be seen that there is no correlation between both quantities, implying that the O/H value by itself is not responsible for the derived \tc~values.

Photoionization models of \Hiirs\ computed with \begin{sc}Cloudy\end{sc} \citep{fer98} find typical \tc~values in the 0.002 to 0.006 range. On the other hand the \tc~values presented in Table \ref{tWMR}~are in the 0.019 to 0.120 range. This difference implies that in addition to photoionization by O stars other physical processes need to be considered in order to explain the observed \tc~values.

\begin{figure}
\begin{center}
\includegraphics[angle=0,scale=0.80]{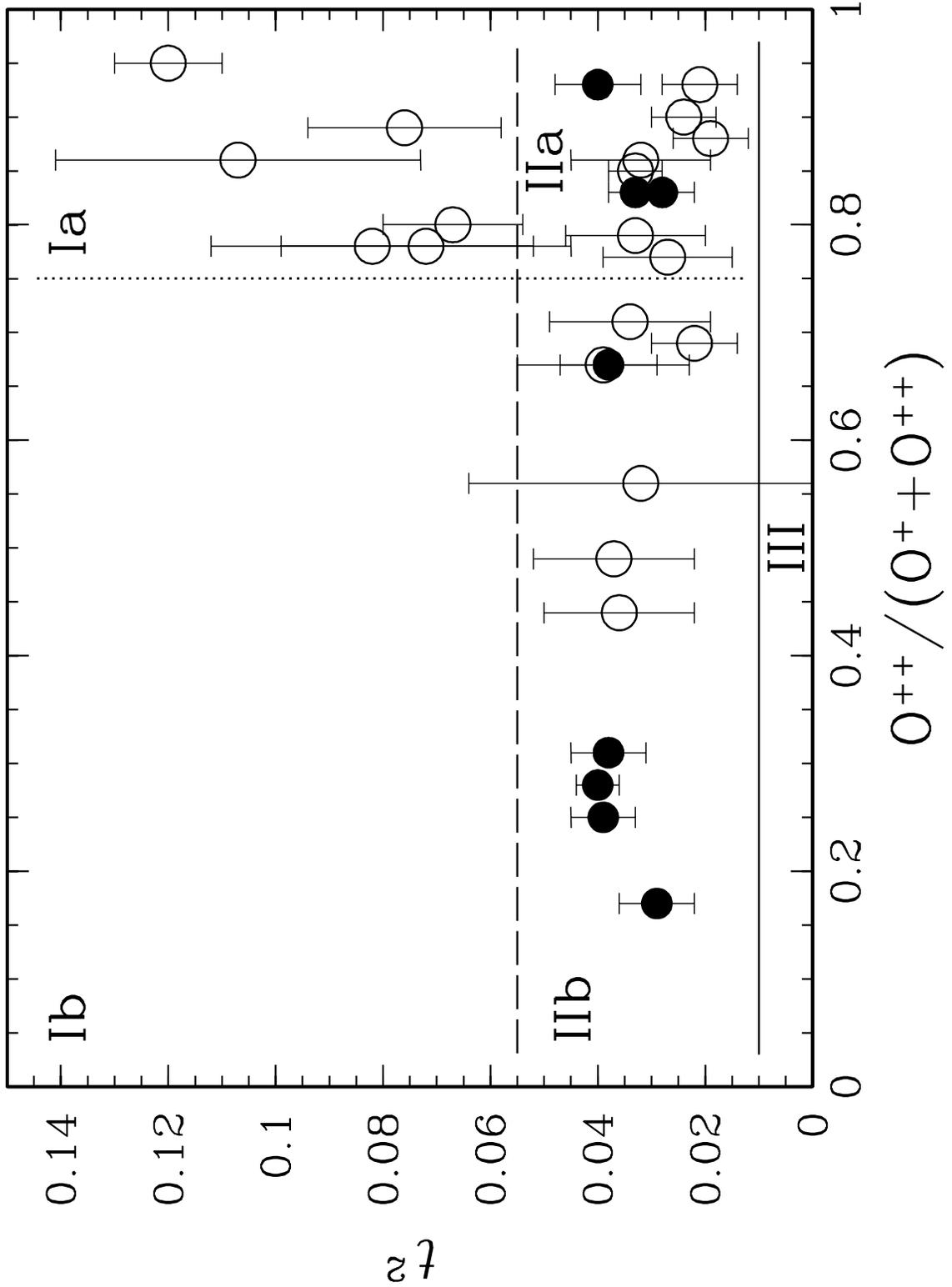}
\caption[f4.eps]{
Plot of ionization degree versus the measured \tc~values. The symbols and error bars are the same as in figure \ref{t2prom}.\label{t2promO}}
\end{center}
\end{figure}

In Figure \ref{t2promO}~we present the \oid\ versus \tc~diagram for the \Hiirs~listed in Table \ref{tWMR}; symbols and errors are the same as those presented in Figure \ref{t2prom}. The objects are not distributed randomly, we therefore divided the figure in five areas: Area Ia (\tc$\ge$0.055, \oid$\ge$0.75), Area Ib (\tc$\ge$0.055, \oid$<$0.75), Area IIa (0.010$<$\tc$<$0.055, \oid$\ge$0.75), Area IIb (0.010$<$\tc$<$0.055, \oid$<$0.75), and Area III (0.000$\le$\tc$\le$0.010). Of the 21 Type II \Hiirs, 8 are Galactic and 13 extragalactic, and of the Galactic ones 5 are of Type IIb and 3 of Type IIa. Notice there are no \Hiirs~in Areas Ib and III in Figure \ref{t2promO}.

We have decided to call the \Hiirs~in Areas Ia, IIa, and IIb as \Hiirs~of Type Ia, Type IIa and Type IIb, respectively. We consider that this division has physical meaning and can be useful for the study of \Hiirs. For Galactic and nearby objects, the observed data represents only a fraction of the object, hence their classification corresponds to the physical conditions of the observed fraction, which could be different from the classification of the whole object. If the observed fraction is close to an O, WR star, or SNR this might affect its classification as well, and viceversa. For example, with the used data we classified the Orion Nebula as a Type IIa but if we were observing the whole object it would be a Type IIb. In what follows we give a general description of these three Types. 

Table \ref{tWMR}~includes six Type Ia \Hiirs: NGC 5253 H II-2, NGC 6822-V, NGC 456, NGC 5471, NGC 2363, and \told. These objects have in common some of the following characteristics: (i) more than a few O stars, (ii) one or more bursts of star formation that have lasted more than three to four million years, (iii) the presence of a relatively large WR to O stars ratio, (iv) in some cases the presence of one or a few SNRs with ages smaller than about twenty thousand years, and (v) a significant spread of gaseous radial motions, reaching in some cases hundreds of km s$^{-1}$ \citep[e.g.][and references therein]{pei91,lur99,che02,lop07,lag11}. Many of these characteristics are related. It would seem that for all of these objects there is a strong contribution from either WR or SNR. It would also seem that they have to have several O stars because one single O star in the presence of a WR or a SNR would be regarded as a WR nebula or a SNR. 

Table \ref{tWMR}~also includes 10 Type IIa and 11 Type IIb \Hiirs. Typically these objects have in common some of the following characteristics: (i) there is no limitation on the number of O stars, (ii) for most of the \Hiirs, the star formation activity of O stars started less than three million years ago, (iii) for the objects older than three million years, they have a small ratio of WR to O stars (at least in the observed fraction of the object), (iv) they do not show vestiges of recent SN explosions (at least in the observed fraction of the object), and (v) their bulk gaseous radial motions show a spread smaller than about 100 km s$^{-1}$.

For observations that are representative of the whole object, the lower degree of ionization of the Type IIb relative to that of the Type IIa \Hiirs\ is due to the later type of their most massive stars. This correlates with a smaller number of O stars formed inside Type IIb \Hiirs. For objects where only a small fraction of their volume is observed, the degree of ionization also depends on how close the observations are to the ionizing stars.

The \tc~values in \Hiirs~are real and have been corroborated by different types of determinations \citep[for a review see][]{pei11}. The lack of Type III \Hiirs~implies that, in addition to photoionization from O stars, additional processes affecting the temperature structure of the nebulae have to be taken into account.

Shadowed regions behind clumps close to the ionization front, can make a significant contribution to the observed temperature inhomogeneities in the Orion nebula \citep{ode03}, which is a Type II \Hiir. It is possible that an additional contribution to the observed temperature inhomogeneities for Type Ia \Hiirs~might be due to the mass loss processes associated with the formation and evolution of O stars.

It has been estimated that the amount of kinetic energy deposited in the interstellar medium (ISM) due to either mass loss by an O star, a WR star, and a Type II SN is about 1-2$\times$10$^{51}$ ergs. The time span for the injection of energy into the ISM by these processes is about 2 to 3 million years for an O star, 200 to 300 thousand years for a WR star, and 20 to 30 thousand years for a SN remnant \citep[e.g.][and references therein]{pei91}. From the time scale of these processes a higher effect on the \tc~values is expected from those objects that at present include WR stars and SNRs. We propose that this is the main reason for the extremely high \tc~values in Type Ia \Hiirs.

In an \Hiir, the effect of shocks on the intensity of [\ion{O}{3}] $\lambda$4363 is very strong, while on the intensity of [\ion{O}{3}] $\lambda$5007 and  [\ion{O}{2}] $\lambda$3727 the effect of shocks is almost negligible \citep[e.g.][]{pei91}. This result implies that, if \tc~is not known and if the RLs of O II have not been detected, it is considerably better to obtain the O/H value by using a photoionization model like \begin{sc}Cloudy\end{sc}~to fit $\lambda \lambda$5007 and 3727 without fitting $\lambda$4363, than to derive the O/H value by using the temperature derived from the 4363/5007 ratio, the so called direct determination. The use of the 4363/5007 ratio under the assumption of \tc=0.00 will underestimate the O/H abundance. For further discussion of this point see for example \citet{mcg91}, \citet{kew08}, and \citet{pen12}.

Even though the \tc~value depends on the specific characteristics of each object, we find that \tc~values for \Hiirs~in Areas IIa and IIb show a very small dispersion. For Type IIa \Hiirs~\tcp$\,=0.029\pm0.002$ with a dispersion of $\sigma(t^2)=0.004$; while for Type IIb \Hiirs~\tcp$\,=0.035\pm0.002$ with $\sigma(t^2)=0.004$. The very small dispersion of Type II \Hiirs, implies that there has to be a series of physical processes present in all \Hiirs, that will produce a typical \tc~value of about 0.030 (considerably larger than those predicted by conventional photoionization models). 

For Type Ia \Hiirs~\tcp$\,=0.087\pm0.009$ with $\sigma(t^2)=0.014$. The large \tc~values of Type Ia regions show that there should be additional physical processes that are less general than those present in Type II \Hiirs; moreover, the large dispersion shows that they must have an stochastic nature. The absence of Type~Ib \Hiirs\ indicates that these processes are associated only with objects of high degree of ionization.

When the observations are not deep/accurate enough to determine \tc, the \oid\ ratio, together with Figure \ref{t2promO}, can be used to estimate the \tc~value of a particular region. When \oid$<0.75$, we are talking about an object of Area IIb and we can assume \tc(IIb)$=0.035\pm0.006$. If \oid$\ge0.75$, it is useful to be able to determine whether the object belongs to Area Ia or IIa, so we can use \tc(Ia)$=0.087\pm0.015$ or \tc(IIa)$=0.029\pm0.004$; without this information we must group all these objects in the same bin and assume that \tc(Ia+IIa)$=0.051\pm0.026$.

\section{Summary and Conclusions}\label{conc}
We determined the physical conditions of \told\ and \tolc, through a detailed analysis involving abundance determinations considering the presence of thermal inhomogeneities. In order to compare with other studies, we also determined abundances assuming a homogeneous thermal structure throughout both objects. We used two methods to determine values of \tc: (i) with a set of \ion{He}{1} lines and the \hecs~program, and (ii) by comparing the abundances obtained from RLs of \ion{O}{2} to the abundances obtained from CELs of [\ion{O}{3}]; we then combined \tc(He$^+$) with \tc(O$^{++}$) and obtained a final value of the thermal inhomogeneities. For \told\ we found \tc=0.091$\pm$0.028 and \tc=0.107$\pm$0.034 for extractions C and E, respectively; for \tolc\ the erros in the \tc\ determination were very large and consequently we adopted \tc(Ia+IIa)$=0.051\pm0.026$, the average value of \Hiirs\ Types Ia and IIa. In addition, for all \Hiirs, total oxygen abundances have to include the fraction of oxygen depleted into dust grains. For \told, both corrections add up to 0.23 dex and 0.28 dex for Extractions C and E, respectively; while for \tolc, they add up to 0.20 dex and 0.21 dex for Extractions C and E, respectively.

Since RLs of heavy elements are too faint in low metallicity objects, one alternative to determine the value of \tc~and its effect on the abundances is through the comparison of the temperatures determined with \ion{He}{1} RLs versus the temperatures obtained using CELs. This can be done by obtaining precise measurements of \ion{He}{1} lines, which are two or three orders of magnitude brighter than oxygen RLs. 

In order to determine values of \tc(He$^+$) with an uncertainty of about 0.015$\,$-$\,$0.020, \hecs~requires about 10 \hels~with S/N greater than 15 (the tenth brightest optical \ion{He}{1} line has an intensity of approximately 1\%~of H$\beta$), and a S/N of about 50 for the brightest \ion{He}{1} line (the brightest optical \ion{He}{1} line is $\lambda$ 5876, which has an intensity of approximately 10\%~of H$\beta$).

Based on the detailed studies of \Hiirs~with a wide metallicity range, the measured \tc~values usually lie between 0.020 and 0.120. This range comes from the thermal structure of each region and not from errors in the determinations. 

For many Galactic and extragalactic \Hiirs~available in the literature neither the oxygen RLs nor the \ion{He}{1} lines have good enough S/N to provide a precise value of \tc; moreover, it is not possible to use any other method to obtain an accurate value of \tc. For these type of objects or observations we propose to use one of the available \tcp, the one that represents the type of the observed \Hiir: Type Ia, Type IIa, or Type IIb. From the sample of objects used in this paper, the average for each type of region is: \tc(Ia)$=0.087\pm0.015$, \tc(IIa)$=0.029\pm0.004$, and \tc(IIb)$=0.035\pm0.006$; for objects of high degreee of ionization without additional information, we recommend \tc(Ia+IIa)$=0.051\pm0.026$.

We find that it is better to use an average value of \tc~than to assume a homogeneous temperature structure; moreover, abundances must be corrected for depletion of oxygen into dust grains. Ignoring these corrections when determining abundances for \Hiirs, will lead to systematically underestimate their abundances by a factor of about 2.

We are grateful to the anonymous referee for a careful reading of the manuscript and many excellent suggestions. 
We are also grateful to Mar\'ia Teresa Ruiz for her assistance in the initial aspects of this work.
AP and MAPG received partial support from UNAM (grant PAPIIT 112911) and AP and MP received partial support from CONACyT (grant 129753).

\clearpage

\begin{deluxetable}{lcr@{--}lcc} 
\tablecaption{Observations Settings
\label{tobs}}
\tablewidth{0pt}
\tablehead{
\colhead{Grism}  & 
\colhead{Filter} &
\multicolumn{2}{c}{$\lambda$ (\AA)} &
\colhead{Resolution ($\lambda$/$\Delta \lambda$)} &
\colhead{Exp. Time (s)}}
\startdata
GRIS-600B+12   &   -   	 & 3450 &5900  & 1300 & 3$\times$720  \\
GRIS-600R+14  & GG435 & 5350 &7450  & 1700 & 3$\times$600  \\
GRIS-300V    	  & GG375 & 3850 &8800  &   700 & 3$\times$120  \\
\enddata
\end{deluxetable} 
\clearpage

\begin{deluxetable}{lcr@{}lr@{}lr@{}lccr@{}lr@{}lccr@{}lr@{}lc}
\tabletypesize{\small}
\tablewidth{0pt}
\tablecaption{\told: Line Intensities
\label{tlines}}
\tablehead{
&&&& \multicolumn{5}{c}{Extraction C} &&\multicolumn{5}{c}{Extraction E} \\
     \cline{5-9}                                       \cline{11-15}\\
\colhead{$\lambda$ (\AA)} & \colhead{ID} & \multicolumn{2}{c}{$f$($\lambda$)} &
\multicolumn{2}{c}{$F(\lambda)$} & \multicolumn{2}{c}{$I(\lambda)$} & \colhead{\%~err} &&
\multicolumn{2}{c}{$F(\lambda)$} & \multicolumn{2}{c}{$I(\lambda)$} & \colhead{\%~err}
}
\startdata

3634    &\ion{He}{1}  &0.&280&    0.&93&   1.&11&   7&&   0.&82&   0.&98& 10\\
3646    &Balmer Cont&0.&278&    0.&13\tablenotemark{a}&   0.&15\tablenotemark{b}& 19&&
											      0.&13\tablenotemark{a}&   0.&16\tablenotemark{b}& 26\\
3676    &H\,22          &0.&269&    0.&39&   0.&52& 11&&  \mcnd &  \mcnd &\nd\\
3679    &H\,21          &0.&268&    0.&24&   0.&36& 14&&  \mcnd &  \mcnd &\nd\\
3683    &H\,20          &0.&267&    0.&33&   0.&50& 12&&   0.&39&   0.&56& 15\\
3687    &H\,19          &0.&266&    0.&54&   0.&76&   9&&   0.&65&   0.&90& 12\\
3692    &H\,18          &0.&265&    0.&57&   0.&84&   9&&   0.&50&   0.&76& 13\\
3695    &H\,17          &0.&264&    0.&86&   1.&22&   7&&   0.&96&   1.&34&9.5\\
3704    &H\,16          &0.&262&    1.&23&   1.&71&  6&&   1.&29&   1.&79&8.5\\
3712    &H\,15          &0.&260&    1.&20&   1.&74&  6&&   1.&25&   1.&80& 8.5\\
3726    &[\ion{O}{2}]  &0.&256&  21.&76& 25.&37&  2&&  22.&82&  26.&60&  2\\
3729    &[\ion{O}{2}]  &0.&255&  31.&04& 36.&41&1.5&&  32.&31&  37.&88& 2\\
3750    &H\,12          &0.&250&    2.&17&   3.&17&7.5&&   2.&22&   3.&24&6.5\\
3770    &H\,11          &0.&245&    2.&76&   4.&03&4.5&&   2.&77&   4.&06&  6\\
3798    &H\,10          &0.&238&    3.&75&   5.&37&3.5&&   3.&79&   5.&42&  5\\
3820    &\ion{He}{1}  &0.&233&    0.&54&   0.&77&  9&&   0.&75&   1.&01& 11\\
3836    &H\,9            &0.&229&    5.&45&   7.&44&  3&&   5.&43&   7.&48&  4\\
3869    &[\ion{Ne}{3}] &0.&222& 35.&26& 40.&46&1.5&&  35.&44&  40.&66&  2\\
3889    &H\,8+\ion{He}{1}
                       		  &0.&218&  14.&88& 18.&59&  2&&  15.&04&  18.&81&2.5\\
3967    &H\,7+\ion{He}{1}+[\ion{Ne}{3}]
		       		  &0.&201&  23.&82&  27.&33&  2&&  24.&17&  27.&72&2\\
4026    &\ion{He}{1}  &0.&190&    1.&41&   1.&83&5.5&&   1.&32&   1.&74&  8\\
4069    &[\ion{S}{2}]  &0.&182&    0.&81&   0.&90&7.5&&   0.&72&   0.&81& 11\\
4076    &[\ion{S}{2}]  &0.&181&    0.&23&   0.&26& 14&&   0.&25&   0.&27& 19\\
4102    &H$\delta$   &0.&176&   21.&86& 25.&95&  2&&  21.&91&  25.&74&  2\\
4121    &\ion{He}{1}  &0.&173&    0.&21&   0.&35& 14&&   0.&22&   0.&36& 20\\
4144    &\ion{He}{1}  &0.&169&    0.&29&   0.&45& 12&&   0.&43&   0.&60& 14\\
4267    &\ion{C}{2}    &0.&143&    0.&04&   0.&04& 33&&   0.&08&   0.&09& 33\\
4340    &H$\gamma$&0.&128&   42.&26&  47.&08&1.5&&  42.&36&  46.&86&  2\\
4363    &[\ion{O}{3}]  &0.&122&   11.&87&  12.&75&  2&&  11.&90&  12.&77&  3\\
4388    &\ion{He}{1}  &0.&116&    0.&36&   0.&39& 11&&   0.&34&   0.&37& 16\\
4471    &\ion{He}{1}  &0.&094&    3.&29&   3.&66&  4&&   3.&31&   3.&67&  5\\
4563    &\ion{Mg}{1}] &0.&070&    0.&23&   0.&24& 14&&  \mcnd &  \mcnd &\nd\\
4571    &\ion{Mg}{1}] &0.&068&    0.&11&   0.&12& 20&&  \mcnd &  \mcnd &\nd\\
4639+42 &\ion{O}{2} &0.&051&    0.&03&   0.&04& 36&&   0.&05&   0.&05& 41\\
4649+51 &\ion{O}{2} &0.&049&    0.&04&   0.&04& 35&&   0.&03&   0.&03& 52\\
4658    &[\ion{Fe}{3}]  &0.&047&    0.&57&   0.&58&  9&&   0.&52&   0.&53& 13\\
4686    &\ion{He}{2}   &0.&041&    1.&69&   1.&72&  5&&   1.&79&   1.&83&  7\\
4701    &[\ion{Fe}{3}]  &0.&037&    0.&08&   0.&08& 24&&  \mcnd &  \mcnd &\nd\\
4711    &[\ion{Ar}{4}]+\ion{He}{1}
	                          &0.&034&    1.&97&   2.&14&  5&&   2.&03&   2.&19&  7\\
4740    &[\ion{Ar}{4}] &0.&028&    1.&19&   1.&20&  6&&   1.&25&   1.&26&8.5\\
4861    &H$\beta$    &0.&000&  100.&00& 100.&00&1&& 100.&00& 100.&00&1.5\\
4881    &[\ion{Fe}{3}] &-0.&004&   0.&20&   0.&20& 15&&   0.&17&   0.&16& 24\\
4922    &\ion{He}{1}  &-0.&013&   1.&06&   1.&13&6.5&&   1.&09&   1.&17&  9\\
4959    &[\ion{O}{3}]  &-0.&021& 201.&88& 197.&33&  1&& 200.&43& 195.&83&  2\\
4986+87&[\ion{Fe}{3}] &-0.&027&   0.&97&   0.&95&  7&&   1.&14&   1.&11&  9\\
5007    &[\ion{O}{3}]  &-0.&032& 612.&30& 594.&14&1&& 608.&01& 589.&76&  1\\
5016    &\ion{He}{1}  &-0.&034&   2.&42&   2.&44&   4&&   2.&61&   2.&62&  6\\
5041    &\ion{Si}{2}    &-0.&040&   0.&21&   0.&21& 14&&   0.&28&   0.&27& 18\\
5048    &\ion{He}{1}   &-0.&041&   0.&12&   0.&21& 19&&  \mcnd &  \mcnd &\nd\\
5159    &[\ion{Fe}{2}] &-0.&065&   0.&09&   0.&09& 22&&  \mcnd &  \mcnd &\nd\\
5192    &[\ion{Ar}{3}] &-0.&072&   0.&12&   0.&11& 19&&  \mcnd &  \mcnd &\nd\\
5198    &[\ion{N}{1}]  &-0.&074&   0.&37&   0.&35& 11&&   0.&33&   0.&31& 16\\
5262    &[\ion{Fe}{3}] &-0.&087&   0.&15&   0.&15& 17&&  \mcnd &  \mcnd &\nd\\
5270    &[\ion{Fe}{3}] &-0.&089&   0.&24&   0.&22& 14&&  \mcnd &  \mcnd &\nd\\
5412    &\ion{He}{2}+[\ion{Fe}{3}]
				  &-0.&119&   0.&14&   0.&12& 18&&  \mcnd &  \mcnd &\nd\\
5517    &[\ion{Cl}{3}] &-0.&140&   0.&30&   0.&28& 12&&   0.&33&   0.&30& 16\\
5537    &[\ion{Cl}{3}] &-0.&144&   0.&13&   0.&12& 18&&   0.&15&   0.&14& 24\\
5876    &\ion{He}{1}  &-0.&216&  12.&59&  10.&91&  2&&  12.&29&  10.&64&  3\\
6300    &[\ion{O}{1}]  &-0.&285&   2.&45&   2.&01&4.5&&   2.&49&   2.&04&  6\\
6312    &[\ion{S}{3}]   &-0.&286&   1.&75&   1.&44&  5&&   1.&74&   1.&43&  7\\
6364    &[\ion{O}{1}]  &-0.&294&   0.&79&   0.&64&7.5&&   0.&79&   0.&64&10.5\\
6371    &\ion{Si}{2}   &-0.&295&   0.&13&   0.&11& 18&&   0.&14&   0.&11& 25\\
6548    &[\ion{N}{2}]  &-0.&320&   1.&21&   0.&97&  6&&   1.&34&   1.&07&  8\\
6563    &H$\alpha$   &-0.&322& 354.&22& 282.&33&  1&& 347.&69& 277.&02&  1\\
6583    &[\ion{N}{2}]  &-0.&324&   3.&43&   2.&75&  4&&   3.&61&   2.&88&  5\\
6678    &\ion{He}{1}  &-0.&337&   2.&62&   2.&11&  4&&   2.&57&   2.&08&  6\\
6716    &[\ion{S}{2}]  &-0.&342&   8.&38&   6.&63&2.5&&   8.&59&   6.&79&3.5\\
6731    &[\ion{S}{2}]  &-0.&343&   6.&62&   5.&23&  3&&   6.&63&   5.&24&  4\\
7065    &\ion{He}{1}  &-0.&383&   3.&80&   2.&95&3.5&&   4.&00&   3.&10&  5\\
7136    &[\ion{Ar}{3}] &-0.&391&   5.&74&   4.&39&  3&&   5.&91&   4.&52&  4\\
7281    &\ion{He}{1}   &-0.&406&   0.&60&   0.&46&  9&&   0.&80&   0.&61&10.5\\
7320    &[\ion{O}{2}]  &-0.&410&   1.&83&   1.&38&  5&&   1.&88&   1.&42&  7\\
7330    &[\ion{O}{2}]  &-0.&411&   1.&34&   1.&02&  6&&   1.&33&   1.&00&  8\\
7751    &[\ion{Ar}{3}] &-0.&452&   1.&43&   1.&05&5.5&&   2.&06&   1.&51&6.5\\
\hline
\multicolumn{2}{l}{$EW$(H$\beta$)\tablenotemark{c}}
&&& \multicolumn{5}{c}{234} && \multicolumn{5}{c}{213}\\
\multicolumn{2}{l}{$EW_{abs}$(H$\beta$)\tablenotemark{c}}
&&& \multicolumn{5}{c}{2.0$\pm$0.1} && \multicolumn{5}{c}{2.1$\pm$0.1}\\
\multicolumn{2}{l}{$C$(H$\beta$)\tablenotemark{d}}
&&& \multicolumn{5}{c}{0.29$\pm$0.03} && \multicolumn{5}{c}{0.26$\pm$0.03}\\
\multicolumn{2}{l}{$F$(H$\beta$) \tablenotemark{e}}
&&& \multicolumn{5}{c}{1.78$\times10^{-14}$} && \multicolumn{5}{c}{1.89$\times10^{-14}$}\\
\multicolumn{2}{l}{$I$(H$\beta$) \tablenotemark{f}}
&&& \multicolumn{5}{c}{3.47$\times10^{-14}$} && \multicolumn{5}{c}{3.69$\times10^{-14}$}\\
\enddata
\tablenotetext{a} {This is the Balmer discontinuity in emission given in units of $F$(H$\beta$)/100 per \AA.}
\tablenotetext{b} {In units of $I$(H$\beta$)/100 per \AA; this is the only hydrogen feature not corrected for the presence of stellar spectra (see text).}
\tablenotetext{c} {Units of $EW$(H$\beta$) and $EW_{abs}$(H$\beta$) are given in \AA.}
\tablenotetext{d} {Units of $C$(H$\beta$) are given in dex.}
\tablenotetext{e} {Units of $F$(H$\beta$) are given in erg s$^{-1}$ cm$^{-2}$.}
\tablenotetext{f} {Units of $I$(H$\beta$) are given in erg s$^{-1}$ cm$^{-2}$.}
\end{deluxetable}
\clearpage

\begin{deluxetable}{lcr@{}lr@{}lr@{}lccr@{}lr@{}lccr@{}lr@{}lc}
\tabletypesize{\small}
\tablewidth{0pt}
\tablecaption{\tolc: Line Intensities
\label{tlines2}}
\tablehead{
&&&& \multicolumn{5}{c}{Extraction C} && \multicolumn{5}{c}{Extraction E} \\
     \cline{5-9}                                        \cline{11-15}\\
\colhead{$\lambda$ (\AA)} & \colhead{ID} & \multicolumn{2}{c}{$f$($\lambda$)} & 
\multicolumn{2}{c}{$F(\lambda)$} & \multicolumn{2}{c}{$I(\lambda)$} & \colhead{\%~err} &&
\multicolumn{2}{c}{$F(\lambda)$} & \multicolumn{2}{c}{$I(\lambda)$} & \colhead{\%~err}
}
\startdata
3634    &\ion{He}{1}   &0.&280&    0.&29&   0.&33& 20&&   0.&20&   0.&22& 37\\
3646    &Balmer Cont &0.&278&    0.&18\tablenotemark{a}&   0.&21\tablenotemark{b}& 25&&
											     0.&21\tablenotemark{a}&   0.&23\tablenotemark{b}& 36\\
3679    &H\,21          &0.&268&    0.&51&   0.&73& 15&&   0.&30&   0.&48& 30\\
3683    &H\,20          &0.&267&    0.&58&   0.&85& 14&&   0.&41&   0.&64& 26\\
3687    &H\,19          &0.&266&    0.&64&   0.&97& 14&&   0.&50&   0.&79& 24\\
3692    &H\,18          &0.&265&    0.&61&   1.&01& 14&&   0.&43&   0.&79& 25\\
3695    &H\,17          &0.&264&    0.&85&   1.&35& 12&&   0.&69&   1.&15& 20\\
3704    &H\,16          &0.&262&    1.&28&   1.&94& 10&&   1.&19&   1.&79& 15\\
3712    &H\,15          &0.&260&    0.&96&   1.&69& 11&&   0.&84&   1.&52& 18\\
3726    &[\ion{O}{2}]  &0.&256&   30.&45&  34.&56&  2&&  34.&44&  37.&37&  3\\
3729    &[\ion{O}{2}]  &0.&255&   40.&10&  46.&67&  2&&  44.&21&  48.&09&  3\\
3750    &H\,12          &0.&250&    1.&97&   3.&25&  8&&   1.&97&   3.&15& 12\\
3770    &H\,11          &0.&245&    2.&46&   4.&10&  7&&   2.&46&   3.&97&10.5\\
3798    &H\,10+[\ion{S}{3}]
	               		   &0.&238&    3.&76&   5.&89&  6&&   3.&80&   5.&72&8.5\\
3820    &\ion{He}{1}   &0.&233&    0.&65&   1.&01& 14&&   0.&84&   1.&17& 18\\
3836    &H\,9             &0.&229&    4.&75&   7.&28&  5&&   5.&13&   7.&39&7.5\\
3869    &[\ion{Ne}{3}] &0.&222&   37.&34&  41.&98&  2&&  41.&08&  44.&29&  3\\
3889    &H\,8+\ion{He}{1}
                       		   &0.&218&   14.&41&  18.&24&  3&&  15.&70&  18.&96&4.5\\
3967    &H\,7+\ion{He}{1}+[\ion{Ne}{3}] 
		       		   &0.&201&   23.&82&  28.&58&2.5&&  25.&69&  29.&54&3.5\\
4009    &\ion{He}{1}   &0.&193&   \mcnd &  \mcnd &\nd&&   0.&17&   0.&41& 40\\
4026    &\ion{He}{1}   &0.&190&    1.&10&   1.&66&10.5&&   1.&12&   1.&64& 16\\
4069    &[\ion{S}{2}]   &0.&182&    0.&93&   1.&03& 11&&   0.&86&   0.&92&  18\\
4076    &[\ion{S}{2}]   &0.&181&    0.&32&   0.&35& 19&&   0.&32&   0.&34& 29\\
4102    &H$\delta$    &0.&176&   21.&42&  25.&68\TC&2.5&&  22.&52&  26.&02\TC&3.5\\
4267    &\ion{C}{2}     &0.&143&    0.&08&   0.&09& 39&&   0.&09&   0.&09& 56\\
4287    &[\ion{Fe}{3}]  &0.&139&   \mcnd &  \mcnd &\nd&&   0.&28&   0.&30& 31\\
4340    &H$\gamma$ &0.&128&   40.&65&  45.&29\TC&  2&&  42.&98&  46.&41\TC&  3\\
4363    &[\ion{O}{3}]   &0.&122&   11.&39&  12.&31&3.5&&  11.&14&  11.&69&  5\\
4471    &\ion{He}{1}    &0.&094&    3.&19&   3.&74&  6&&   3.&50&   3.&97&  9\\
4639+42 &\ion{O}{2}  &0.&051&    0.&10&   0.&11& 34&&  \mcnd &  \mcnd &\nd\\
4658    &[\ion{Fe}{3}]  &0.&047&    0.&59&   0.&62& 14&&   0.&90&   0.&92& 18\\
4686    &\ion{He}{2}   &0.&041&    1.&89&   1.&98&  8&&   1.&33&   1.&36&  14\\
4711    &[\ion{Ar}{4}]+\ion{He}{1}
                       		   &0.&034&    1.&69&   2.&00&  8&&   1.&86&   2.&14& 12\\
4740    &[\ion{Ar}{4}]  &0.&028&    1.&01&   1.&05& 11&&   0.&98&   1.&01& 17\\
4861    &H$\beta$      &0.&000&  100.&00& 102.&20\TC&1.5&& 100.&00& 100.&80\TC&  2\\
4922    &\ion{He}{1}   &-0.&013&   0.&83&   1.&01& 12&&   0.&97&   1.&14& 17\\
4959    &[\ion{O}{3}]  &-0.&021& 211.&98& 216.&34&1.5&& 208.&46& 210.&39& 1.5\\
4986+87&[\ion{Fe}{3}] &-0.&027&   1.&02&   1.&04& 11&&   1.&30&   1.&31& 15\\
5007    &[\ion{O}{3}]  &-0.&032& 644.&58& 654.&95&  1&& 626.&16& 630.&08&  1\\
5016    &\ion{He}{1}   &-0.&034&   2.&34&   2.&62&  7&&   2.&44&   2.&68& 11\\
5026    &\ion{N}{2}    &-0.&036&  \mcnd &  \mcnd &\nd&&   0.&25&   0.&26& 33\\
5032    &\ion{S}{2}+[\ion{Fe}{3}]
    		       		  &-0.&038&  \mcnd &  \mcnd &\nd&&   0.&26&   0.&26& 32\\
5041    &\ion{Si}{2}    &-0.&040&   0.&65&   0.&66& 14&&   0.&53&   0.&54& 23\\
5048    &\ion{He}{1}   &-0.&041&   0.&55&   0.&79& 15&&   0.&21&   0.&43& 36\\
5198    &[\ion{N}{1}]   &-0.&074&   0.&25&   0.&25& 22&&   0.&50&   0.&50& 23\\
5262    &[\ion{Fe}{2}] &-0.&087&   0.&31&   0.&31& 19&&  \mcnd &  \mcnd &\nd\\
5270    &[\ion{Fe}{3}] &-0.&089&   0.&22&   0.&22& 23&&  \mcnd &  \mcnd &\nd\\
5517    &[\ion{Cl}{3}]  &-0.&140&   0.&33&   0.&32& 19&&   0.&30&   0.&29& 30\\
5537    &[\ion{Cl}{3}]  &-0.&144&   0.&25&   0.&24& 22&&$<$0.&30&$<$0.&30&\nd\\
5876    &\ion{He}{1}   &-0.&216&  11.&61&  11.&17&  3&& 11.&04&  10.&75& 5\\
6300    &[\ion{O}{1}]   &-0.&285&   2.&52&   2.&31&  7&&   2.&65&   2.&49& 10\\
6312    &[\ion{S}{3}]   &-0.&286&   1.&33&   1.&22&9.5&&   1.&20&   1.&12& 15\\
6364    &[\ion{O}{1}]  &-0.&294&   0.&76&   0.&70& 13&&   0.&78&   0.&73& 19\\
6548    &[\ion{N}{2}]   &-0.&320&   2.&18&   1.&98&7.5&&   2.&24&   2.&01& 11\\
6563    &H$\alpha$    &-0.&322& 319.&29& 280.&80\TC&  1&& 311.&15& 281.&06\TC&1.5\\
6583    &[\ion{N}{2}]    &-0.&324&   6.&18&   5.&59&4.5&&   6.&26&   5.&82&  7\\
6678    &\ion{He}{1}    &-0.&337&   2.&96&   2.&75&6.5&&   2.&92&   2.&78& 10\\
6716    &[\ion{S}{2}]    &-0.&342&   7.&26&   6.&52&  4&&   7.&64&   7.&07&  6\\
6731    &[\ion{S}{2}]    &-0.&343&   5.&56&   4.&99&  5&&   5.&93&   5.&48&  7\\
7065    &\ion{He}{1}    &-0.&383&   3.&99&   3.&56&5.5&&   4.&00&   3.&71&8.5\\
7136    &[\ion{Ar}{3}]  &-0.&391&   4.&61&   4.&06&  6&&   4.&48&   4.&09&  8\\
7281    &\ion{He}{1}    &-0.&406&   0.&48&   0.&45& 16&&   1.&05&   0.&98& 16\\
7320    &[\ion{O}{2}]    &-0.&410&   1.&51&   1.&32&  9&&   2.&13&   1.&93& 11\\
7330    &[\ion{O}{2}]    &-0.&411&   1.&26&   1.&10& 10&&   1.&53&   1.&39& 13\\
7751    &[\ion{Ar}{3}]   &-0.&452&   2.&56&   2.&20&  7&&   2.&20&   1.&98& 11\\

\hline
\multicolumn{2}{l}{$EW$(H$\beta$)\tablenotemark{d}}
&&& \multicolumn{5}{c}{236} && \multicolumn{5}{c}{196}\\
\multicolumn{2}{l}{$EW_{abs}$(H$\beta$)\tablenotemark{d}}
&&& \multicolumn{5}{c}{3.5$\pm$0.1} && \multicolumn{5}{c}{2.9$\pm$0.1}\\
\multicolumn{2}{l}{$C$(H$\beta$)\tablenotemark{e}}
&&& \multicolumn{5}{c}{0.17$\pm$0.03} && \multicolumn{5}{c}{0.12$\pm$0.03}\\
\multicolumn{2}{l}{$F$(H$\beta$) \tablenotemark{f}}
&&& \multicolumn{5}{c}{1.34$\times10^{-14}$} && \multicolumn{5}{c}{1.51$\times10^{-14}$}\\
\multicolumn{2}{l}{$I$(H$\beta$) \tablenotemark{g}}
&&& \multicolumn{5}{c}{2.03$\times10^{-14}$} && \multicolumn{5}{c}{2.01$\times10^{-14}$}\\
\enddata
\tablenotetext{a} {This is the Balmer discontinuity in emission given in units of $F$(H$\beta$)/100 per \AA.}
\tablenotetext{b} {In units of $I$(H$\beta$)/100 per \AA; this is the only hydrogen feature not corrected for the presence of stellar spectra (see text).}
\tablenotetext{c} {In units of the best fit to the theoretical Balmer decrement.}
\tablenotetext{d} {Units of $EW$(H$\beta$) and $EW_{abs}$(H$\beta$) are given in \AA.}
\tablenotetext{e} {Units of $C$(H$\beta$) are given in dex.}
\tablenotetext{f} {Units of $F$(H$\beta$) are given in erg s$^{-1}$ cm$^{-2}$.}
\tablenotetext{g} {Units of $I$(H$\beta$) are given in erg s$^{-1}$ cm$^{-2}$.}
\end{deluxetable}
\clearpage

\begin{deluxetable}{ccccc}
\tablecaption{Absorption Equivalent Widths Relative to EW$_{abs}$(H$\beta$)
\label{tEWs}}
\tablewidth{0pt}
\tablehead{
\multicolumn{2}{c}{Hydrogen} && \multicolumn{2}{c}{Helium} \\
\cline{1-2}  \cline{4-5}
\colhead{Line} & \colhead{$\frac{\rm EW_{abs}(\rm Line) }{\rm EW_{abs}(\rm H\beta)}$}
&& \colhead{Line} & \colhead{$\frac{\rm EW_{abs}(\rm Line) }{\rm EW_{abs}(\rm H\beta)}$}
}
\startdata

H$\alpha$\tablenotemark{a}	&0.90	&& 3820\tablenotemark{b}	&0.108\\
H$\gamma$\tablenotemark{b}&1.05	&& 4009\tablenotemark{a}	&0.100\\
H$\delta$\tablenotemark{b}	&1.08	&& 4026\tablenotemark{b}	&0.193\\
H7\tablenotemark{a}		&0.99	&& 4121\tablenotemark{a}	&0.094\\
H8\tablenotemark{b,c}		&0.93	&& 4143\tablenotemark{a}	&0.104\\
H9\tablenotemark{b}		&0.78	&& 4388\tablenotemark{b}	&0.084\\
H10\tablenotemark{b}		&0.67	&& 4471\tablenotemark{b}	&0.179\\
H11\tablenotemark{d}		&0.54	&& 4922\tablenotemark{b}	&0.107\\
H12\tablenotemark{d}		&0.42	&& 5016\tablenotemark{a}	&0.114\\
H13\tablenotemark{d}		&0.35	&& 5048\tablenotemark{a}	&0.115\\
H14\tablenotemark{d}		&0.30	&& 5876\tablenotemark{a}	&0.138\\
H15\tablenotemark{d}		&0.25	&& 6678\tablenotemark{a}	&0.082\\
H16\tablenotemark{d}		&0.20	&& 7065\tablenotemark{a}	&0.073\\
H17\tablenotemark{d}		&0.16	&& 7281\tablenotemark{a}	&0.032\\
H18\tablenotemark{d}		&0.13	&& \\
H19\tablenotemark{d}		&0.10	&& \\
H20\tablenotemark{d}		&0.08	&& \\
H21\tablenotemark{d}		&0.06	&& \\
H22\tablenotemark{d}		&0.05	&& \\

\enddata
\tablenotetext{a} {Private communication with M. Cervi\~no, 2006.}
\tablenotetext{b} {From \citet{gon99}.}
\tablenotetext{c} {Note that \ion{He}{1}(3889) is blended with H8, and both are corrected together.}
\tablenotetext{d} {Extrapolations based on 30 Doradus \citep{pea03}. }
\end{deluxetable}
\clearpage

\begin{deluxetable}{lr@{}lr@{}lcr@{}lr@{}l}
\tablecaption{Densities and Temperatures
\label{tdat}}
\tablewidth{0pt}
\tablehead{
 & \multicolumn{4}{c}{\told} && \multicolumn{4}{c}{\tolc} \\
    \cline{2-5}       \cline{7-10}
 & \multicolumn{2}{c}{Extraction C} & \multicolumn{2}{c}{Extraction E}
 && \multicolumn{2}{c}{Extraction C}& \multicolumn{2}{c}{Extraction E}}
\startdata
Densities (cm$^{-3}$)\\
\ion{He}{1}	 &              $390$&$\pm190$&             $660$&$\pm255$&&              $240$&$\pm95$ & $230$&$\pm95$ \\
$[$\ion{O}{2}$]$ &              $280$&$\pm30$ &             $280$&$\pm30$ &&              $340$&$\pm50$ & $350$&$\pm90$ \\
$[$\ion{S}{2}$]$ &              $160$&$\pm30$ &             $125$&$\pm35$ &&              $110$&$\pm50$ & $130$&$\pm75$ \\
$[$\ion{Cl}{3}$]$& \multicolumn{2}{c}{$<2000$}&\multicolumn{2}{c}{$<1000$}&&\multicolumn{2}{c}{$<2000$} &  \mcnd    \\
$[$\ion{Fe}{3}$]$&              $64$&$\pm12$ &                $43$&$\pm16$ &&               $60$&$\pm18$ & $78$&$\pm32$ \\
\\
\hline
\\
Temperatures (K)\\ 
\ion{He}{1}         & $13850$&$\pm750$            & $13300$&$\pm1250$          && $13690$&$\pm1650$                & $14450$&$\pm2000$ \\
$[$\ion{O}{2}$]$ & $14120$&$\pm420$	             & $14140$&$\pm600$             && $13600$&$\pm^{370}_{800}$    & $13620$&$\pm950$  \\
$[$\ion{S}{2}$]$ & $11320$&$\pm^{390}_{850}$& $10100$&$\pm^{850}_{660}$&& $13620$&$\pm^{2050}_{1400}$& $11390$&$\pm^{1975}_{1260}$\\
$[$\ion{O}{3}$]$ & $15800$&$\pm170$            & $15810$&$\pm225$              && $14870$&$\pm230$                  & $14770$&$\pm350$  \\
\enddata
\end{deluxetable}
\clearpage

\begin{deluxetable}{lccccccccccc}
\tabletypesize{\small}
\tablecaption{Ionic Abundance Determinations from CELs\tablenotemark{a}
\label{tceionic}}
\tablewidth{0pt}
\tablehead{
\colhead{Ion}
&\multicolumn{5}{c}{\told} \\
&\multicolumn{2}{c}{Extraction C}   && \multicolumn{2}{c}{Extraction E}\\
 \cline{2-3}                         \cline{5-6}
&\colhead{$t^2 = 0.000$}&\colhead{$t^2 = 0.091\pm0.028$}
&&\colhead{$t^2 = 0.000$}&\colhead{$t^2 = 0.107\pm0.034$}
}
\startdata
N$^+$       & 5.36$\pm$0.06 & 5.63$\pm$0.13 && 5.39$\pm$0.08 & 5.73$\pm$0.17\\
O$^0$       & 6.05$\pm$0.09 & 6.33$\pm$0.03 && 6.06$\pm$0.12 & 6.41$\pm$0.03\\
O$^+$      & 6.83$\pm$0.09 & 7.16$\pm$0.15 && 6.84$\pm$0.13 & 7.25$\pm$0.29\\
O$^{++}$  & 7.74$\pm$0.02 & 7.96$\pm$0.10 && 7.74$\pm$0.03 & 8.00$\pm$0.12\\
Ne$^{++} $& 6.96$\pm$0.03 & 7.20$\pm$0.10 && 6.97$\pm$0.02 & 7.25$\pm$0.13\\
S$^+$       & 5.07$\pm$0.06 & 5.34$\pm$0.12 && 5.07$\pm$0.08 & 5.41$\pm$0.17\\
S$^{++}$  & 5.81$\pm$0.03 & 6.04$\pm$0.10 && 5.81$\pm$0.03 & 6.09$\pm$0.12\\
Cl$^{++}$ & 3.86$\pm$0.09 & 4.07$\pm$0.10 && 3.91$\pm$0.12 & 4.16$\pm$0.13\\
Ar$^{++}$ & 5.17$\pm$0.03 & 5.35$\pm$0.09 && 5.30$\pm$0.03 & 5.53$\pm$0.11\\
Ar$^{+3}$ & 4.93$\pm$0.04 & 5.15$\pm$0.24 && 4.95$\pm$0.06 & 5.22$\pm$0.43\\
\\
\hline
\\
&\multicolumn{5}{c}{\tolc}  \\
&\multicolumn{2}{c}{Extraction C}   && \multicolumn{2}{c}{Extraction E}\\
 \cline{2-3}                         \cline{5-6}
&\colhead{$t^2 = 0.000$}&\colhead{$t^2 = 0.051\pm0.026$}
&&\colhead{$t^2 = 0.000$}&\colhead{$t^2 = 0.051\pm0.026$}
\\
\hline
\\
N$^+$       & 5.69$\pm$0.14 & 5.84$\pm$0.16 && 5.72$\pm$0.14 & 5.86$\pm$0.16\\
O$^0$       & 6.16$\pm$0.15 & 6.30$\pm$0.17 && 6.19$\pm$0.21 & 6.34$\pm$0.23\\
O$^+$       & 6.98$\pm$0.16 & 7.22$\pm$0.18 && 7.02$\pm$0.22 & 7.20$\pm$0.25\\
O$^{++}$   & 7.84$\pm$0.04 & 7.97$\pm$0.10 && 7.84$\pm$0.05 & 7.97$\pm$0.09\\
Ne$^{++}$  & 7.15$\pm$0.03 & 7.25$\pm$0.08 && 7.05$\pm$0.06 & 7.19$\pm$0.09\\
S$^+$        & 5.11$\pm$0.10 & 5.25$\pm$0.12 && 5.13$\pm$0.13 & 5.27$\pm$0.15\\
S$^{++}$   & 5.81$\pm$0.09 & 5.94$\pm$0.11 && 5.79$\pm$0.15 & 5.93$\pm$0.17\\
Cl$^{++}$  & 4.07$\pm$0.14 & 4.19$\pm$0.15 && 3.98$\pm$0.22 & 4.10$\pm$0.23\\
Ar$^{++}$  & 5.18$\pm$0.04 & 5.28$\pm$0.07 && 5.19$\pm$0.07 & 5.30$\pm$0.09\\
Ar$^{+3}$  & 4.92$\pm$0.07 & 4.04$\pm$0.10 && 4.91$\pm$0.11 & 5.04$\pm$0.13\\
\enddata
\tablenotetext{a}{In units of 12 + log$\,n(X^{+i})/n$(H),
 gaseous content only.}
\end{deluxetable}
\clearpage

\begin{deluxetable}{lcccccc}
\tabletypesize{\small}
\tablecaption{Ionic Abundance Determinations from RLs\tablenotemark{a} \label{trionic}}
\tablewidth{0pt}
\tablehead{
\colhead{Ion}
&\multicolumn{5}{c}{\told} \\
& \multicolumn{2}{c}{Extraction C} && \multicolumn{2}{c}{Extraction E}\\
\cline{2-3}
\cline{5-6}
&\colhead{$t^2 = 0.000$}&\colhead{$t^2 = 0.091\pm0.028$}
&&\colhead{$t^2 = 0.000$}&\colhead{$t^2 = 0.107\pm0.034$}
}
\startdata
He$^+$     & 10.849$\pm$0.015 & 10.834$\pm$0.017 && 10.843$\pm$0.019 & 10.833$\pm$0.031 \\
He$^{++}$ &   9.17$\pm$0.02    &   9.16$\pm$0.02     &&    9.20$\pm$0.03    &  9.19$\pm$0.03      \\
O$^{++}$  &   7.94$\pm$0.11     &  7.92$\pm$0.11      &&   8.02$\pm$0.14     &  8.00$\pm$0.14      \\
\\
\hline
\\
&\multicolumn{5}{c}{\tolc}  \\
&\multicolumn{2}{c}{Extraction C}   && \multicolumn{2}{c}{Extraction E}\\
 \cline{2-3}                         \cline{5-6}
&\colhead{$t^2 = 0.000$}&\colhead{$t^2 = 0.051\pm0.026$}
&&\colhead{$t^2 = 0.000$}&\colhead{$t^2 = 0.051\pm0.026$}
\\
\hline
\\
He$^+$    & 10.909$\pm$0.011 & 10.904$\pm$0.011 && 10.914$\pm$0.015 & 10.908$\pm$0.015 \\
He$^{++}$&  9.08$\pm$0.06     &    9.07$\pm$0.06    &&    9.22$\pm$0.03    &    9.20$\pm$0.03     \\
\enddata
\tablenotetext{a} {In units of 12 + log$\,n(X^{+i})/n$(H), gaseous content only.}
\end{deluxetable}
\clearpage

\begin{deluxetable}{lccccccc} 
\tablecaption{Ionization Correction Factors
\label{tICF}}
\tablewidth{0pt}
\tablehead{
\colhead{Element}  & 
\multicolumn{5}{c}{ICF} &
\colhead{Reference} \\
\hline
\\
&\multicolumn{5}{c}{\told} \\
&\multicolumn{2}{c}{Extraction C}   && \multicolumn{2}{c}{Extraction E}\\
 \cline{2-3}                         \cline{5-6}
&\colhead{$t^2 = 0.000$}&\colhead{$t^2 = 0.091\pm0.028$}
&&\colhead{$t^2 = 0.000$}&\colhead{$t^2 = 0.107\pm0.034$}}
\startdata
N    & 9.18 & 7.24 && 8.93 & 6.65 & 1 \\ 
O    & 1.02 & 1.02 && 1.02 & 1.02 & 2 \\
Ne  & 1.12 & 1.16 && 1.13 & 1.18 & 1 \\
S     & 1.26 & 1.26 && 1.26 & 1.26 & 3 \\
Cl   & 1.36 & 1.36 && 1.36 & 1.36 & 4 \\
Ar   & 1.12 & 1.10 && 1.13 & 1.11 & 5 \\
\\
\hline
\\
&\multicolumn{5}{c}{\tolc}  \\
&\multicolumn{2}{c}{Extraction C}   && \multicolumn{2}{c}{Extraction E}\\
 \cline{2-3}                         \cline{5-6}
&\colhead{$t^2 = 0.000$}&\colhead{$t^2 = 0.051\pm0.026$}
&&\colhead{$t^2 = 0.000$}&\colhead{$t^2 = 0.051\pm0.026$}
\\
\hline
\\
N    & 8.19 & 7.24 && 7.58 & 6.65 & 1 \\ 
O    & 1.02 & 1.01 && 1.01 & 1.02 & 2 \\
Ne  & 1.14 & 1.17 && 1.13 & 1.15 & 1 \\
S     & 1.26 & 1.26 && 1.26 & 1.26 & 3 \\
Cl   & 1.36 & 1.36 && 1.36 & 1.36 & 4 \\
Ar   & 1.14 & 1.13 && 1.14 & 1.14 & 5 \\
\enddata
\\
References.--- (1) \citet{pei69}, (2) \citet{tor77}, (3) \citet{pea05}, (4) \citet{gar89}, and (5) \citet{liu00}.
\end{deluxetable} 
\clearpage

\begin{deluxetable}{ccccccc}
\tabletypesize{\small}
\tablecaption{\told: Total Abundance Determinations\tablenotemark{a}
\label{tabund}}
\tablewidth{0pt}
\tablehead{
\colhead{Element}
&\multicolumn{2}{c}{Extraction C} && \multicolumn{2}{c}{Extraction E} \\
\cline{2-3}
\cline{5-6}
& \colhead{$t^2 = 0.000$} & \colhead{$t^2 = 0.091\pm0.028$}
&& \colhead{$t^2 = 0.000$} & \colhead{$t^2 = 0.107\pm0.034$}
}
\startdata
He\tablenotemark{b}    &   10.849$\pm$0.011 & 10.844$\pm$0.017  &&  10.843$\pm$0.019 & 10.833$\pm$0.031  \\
N\tablenotemark{c}      &    6.32 $\pm$0.01  &  6.49 $\pm$0.08   &&   6.34 $\pm$0.02  &  6.55 $\pm$0.10   \\
O\tablenotemark{b,d}     &    8.09 $\pm$0.11  &  8.08 $\pm$0.11   &&   8.17 $\pm$0.14  &  8.17 $\pm$0.14   \\
O\tablenotemark{c,d}  &    7.89 $\pm$0.01  &  8.12 $\pm$0.09   &&   7.89 $\pm$0.03  &  8.17 $\pm$0.11   \\
Ne\tablenotemark{c}    &    7.02$\pm$0.02  &  7.26 $\pm$0.11   &&   7.02 $\pm$0.02  &  7.32 $\pm$0.13   \\
S\tablenotemark{c}      &    5.98 $\pm$0.04  &  6.22 $\pm$0.09   &&   5.98 $\pm$0.04  &  6.28 $\pm$0.11   \\
Cl\tablenotemark{c}    &    4.00 $\pm$0.06  &  4.21 $\pm$0.09   &&   4.04 $\pm$0.07  &  4.30 $\pm$0.11   \\
Ar\tablenotemark{c}    &    5.42 $\pm$0.09  &  5.63 $\pm$0.10   &&   5.51 $\pm$0.09  &  5.77 $\pm$0.11   \\
\enddata
\tablenotetext{a} {In units of 12 + log$\,n(X)/n$(H).}
\tablenotetext{b} {Recombination lines.}
\tablenotetext{c} {Collisionally excited lines.}
\tablenotetext{d}{O abundance has been corrected for the fraction trapped in dust grains, see text.}
\end{deluxetable}
\clearpage

\begin{deluxetable}{ccccccc}
\tabletypesize{\small}
\tablecaption{\tolc: Total Abundance Determinations\tablenotemark{a}
\label{tabund2}}
\tablewidth{0pt}
\tablehead{
\colhead{Element}
&\multicolumn{2}{c}{Extraction C} && \multicolumn{2}{c}{Extraction E} \\
\cline{2-3}
\cline{5-6}
& \colhead{$t^2 = 0.000$} & \colhead{$t^2 = 0.051\pm0.026$}
&& \colhead{$t^2 = 0.000$} & \colhead{$t^2 = 0.051\pm0.026$}
}
\startdata
He\tablenotemark{b}   & 10.916$\pm$0.013 & 10.915$\pm$0.012  &&  10.922$\pm$0.017 & 10.922$\pm$0.017  \\
N\tablenotemark{c}     &    6.60 $\pm$0.02  &  6.70 $\pm$0.11       &&   6.60 $\pm$0.03  &  6.69 $\pm$0.11   \\
O\tablenotemark{c,d}  &    7.90 $\pm$0.02  & 8.03 $\pm$0.12        &&   7.90 $\pm$0.03  &  8.04 $\pm$0.12   \\
Ne\tablenotemark{c}   &    7.21 $\pm$0.03  &  7.35 $\pm$0.13       &&   7.11 $\pm$0.04  &  7.26 $\pm$0.13   \\
S\tablenotemark{c}     &    5.99 $\pm$0.05  & 6.13 $\pm$0.13        &&   5.98 $\pm$0.07  &  6.12 $\pm$0.13   \\
Cl\tablenotemark{c}    &    4.20 $\pm$0.08  & 4.32 $\pm$0.11        &&   4.12 $\pm$0.12  &  4.24 $\pm$0.11   \\
Ar\tablenotemark{c}    &    5.42 $\pm$0.09  & 5.53 $\pm$0.08        &&   5.44 $\pm$0.10  &  5.55 $\pm$0.08   \\
\enddata
\tablenotetext{a} {In units of 12 + log$\,n(X)/n$(H).}
\tablenotetext{b} {Recombination lines.}
\tablenotetext{c} {Collisionally excited lines.}
\tablenotetext{d}{O abundance has been corrected for the fraction trapped in dust grains, see text.}
\end{deluxetable}
\clearpage

\begin{deluxetable}{lccccccccc} 
\rotate
\tabletypesize{\small}
\tablecaption{Compilation of \Hiir\ \tc~values \label{tWMR}}
\tablewidth{0pt}
\tablehead{
\colhead{Object}  & \colhead{Location\tablenotemark{a}}   & \colhead{O/H\tablenotemark{b}}  & 
\colhead{O/H\tablenotemark{c}} & \colhead{\tc}  & \colhead{O$^{++}$}  &\colhead{$T_e$[\ion{O}{3}]} & \colhead{$n_e$[\ion{O}{2}]} & \colhead{References} & \colhead{Type\tablenotemark{d}} \\
										\cline{6-6}
									&&&&&\colhead{O$^+$+O$^{++}$}
}
\startdata
NGC 3576	&	G&	8.56	&	8.92	&	0.038$\pm$0.009	&	0.67	 &8500$\pm$50	&	2300$\pm$200	& 1& IIb\\
M16			&	G&	8.50	&	8.90	&	0.039$\pm$0.006	&	0.25	 &7650$\pm$250&	1050$\pm$250	& 2& IIb\\
M17			&	G&	8.52	&	8.88	&	0.033$\pm$0.005	&	0.83	&8950$\pm$380&	480	$\pm$150	& 3	& IIa \\
M8			&	G&	8.51	&	8.85	&	0.040$\pm$0.004	&	0.28	&8090$\pm$140&	1800$\pm$800	& 3& IIb \\
H1013		&	X&	8.45	&	8.84	&	0.037			&	0.49	 &7370$\pm$630&	280	$\pm$60		& 4& IIb\\
NGC 595		&	X&	8.45	&	8.80	&	0.036			&	0.44	&7450$\pm$330&	260	$\pm$30		& 4& IIb \\
M20			&	G&	8.53	&	8.79	&	0.029$\pm$0.007	&	0.17	 &7800$\pm$300&	240	$\pm$70		& 2& IIb\\
Orion		&	G&	8.51	&	8.79	&	0.028$\pm$0.006	&	0.83	 &8300$\pm$40	&	2400$\pm$300	& 5, 6& IIa\\
NGC 3603	&	G&	8.46	&	8.78	&	0.040$\pm$0.008	&	0.93	&9060$\pm$200&	2300$\pm$750	& 2& IIa \\
K932		&	X&	8.41	&	8.73	&	0.033			&	0.79	 &8360$\pm$150&	470	$\pm$40		& 4& IIa\\
NGC 2403	&	X&	8.36	&	8.72	&	0.039			&	0.67	&8270$\pm$210&	370	$\pm$40		& 4& IIb \\
NGC 604		&	X&	8.38	&	8.71	&	0.034$\pm$0.015	&	0.71	 &8150$\pm$160&	270	$\pm$30		& 4& IIb\\
S 311		&	G&	8.39	&	8.67	&	0.038$\pm$0.007	&	0.31	&9000$\pm$200&	260	$\pm$110	& 7& IIb \\
NGC 5447	&	X&	8.35	&	8.63	&	0.032			&	0.86	 &9280$\pm$180&280$\pm^{690}_{280}$& 4& IIa\\
30 Doradus	&	X&	8.33	&	8.61	&	0.033$\pm$0.005	&	0.85	&9950$\pm$60	&	279	$\pm$16		& 8& IIa \\
NGC 5461	&	X&	8.41	&	8.60	&	0.027$\pm$0.012	&	0.77	&8470$\pm$200&	540	$\pm$110\tablenotemark{e}& 4, 9& IIa \\
NGC 5253	&	X&	8.18	&	8.56	&	0.072$\pm$0.027	&	0.78	 &11960$\pm$290&	660	$\pm$140	& 10& Ia\\
NGC 6822	&	X&	8.08	&	8.45	&	0.076$\pm$0.018	&	0.89	&13000$\pm$1000&	190	$\pm$30		& 11& Ia \\
NGC 5471	&	X&	8.03 &	8.33 &	0.082$\pm$0.030	&	0.78 &14100$\pm$300&	220	$\pm$70\tablenotemark{f}& 9& Ia \\
NGC 456  	&	X&	7.99	&	8.33	&	0.067$\pm$0.013	&	0.80	 &12165$\pm$200&	130	$\pm$30		& 12& Ia\\
NGC 346		&	X&	8.07	&	8.23	&	0.022$\pm$0.008	&	0.69 &13070$\pm$50&	144$\pm^{44}_{38}$\tablenotemark{g}& 13, 14& IIb\\
NGC 460		&	X&	7.96	&	8.19	&	0.032$\pm$0.032	&	0.56	&12400$\pm$450&	170	$\pm$20		& 12& IIb \\
NGC 2363	&	X&	7.76	&	8.14	&	0.120$\pm$0.010	&	0.95	 &16200$\pm$300&	550	$\pm$100	& 4& Ia\\
\told			&	X&	7.79	&	8.09	&	0.107$\pm$0.034	&	0.86	 &15800$\pm$170&	280	$\pm$30		& 15& Ia\\
Haro 29		&	X&	7.87	&	8.05	&	0.019$\pm$0.007	&	0.88 &16050$\pm$100&	235	$\pm$85\tablenotemark{g}& 13, 16& IIa\\
SBS 0335$-$052&	X&	7.35	&	7.60	&	0.021$\pm$0.007	&	0.93 &20500$\pm$200&	297	$\pm$85\tablenotemark{g}& 13, 17& IIa\\
I Zw 18		&	X&	7.22	&	7.41	&	0.024$\pm$0.006	&	0.90&19060$\pm$610&	87$\pm^{65}_{56}$\tablenotemark{g}& 13, 17& IIa \\
\hline
\\
 $\langle$\tc$\rangle=0.044$
\enddata
\tablenotetext{a}{G=Galactic object, X=Extragalactic object.}
\tablenotetext{b}{Total O abundance with homogeneous temperature, \tc=0.000. In units of 12+log(O/H).}
\tablenotetext{c}{Total O abundance with thermal inhomogeneities, \tc$\neq$0.000, plus the correction due to depletion of O into dust grains \citep{pea10}. In units of 12+log(O/H)}
\tablenotetext{d}{The Type of \Hiir\ corresponds to the clasification we present in this work.}
\tablenotetext{e}{Derived from [\ion{Cl}{3}] lines.}
\tablenotetext{f}{Derived from [\ion{S}{2}] lines.}
\tablenotetext{g}{Derived from \hels.}
\\
References.--- (1) \citet{gar04}; (2) \citet{gar06}; (3) \citet{garb07}; (4) \citet{est09}; (5) \citet{est04}; (6) \citet{ode03}; (7) \citet{gar05}; (8) \citet{pea03}; (9) \citet{est02}; (10) \citet{lop07}; (11) \citet{pea05}; (12) \citet{pen12}; (13) \citet{pei07}; (14) \citet{pei00}; (15) This work; (16) \citet{izo97}; (17) \citet{izo99}.
\end{deluxetable} 
\clearpage

\clearpage

\end{document}